\documentclass[a4paper,11pt]{article}
\pdfoutput=1 
\usepackage{jheppub} 
\usepackage[T1]{fontenc} 
\usepackage{multirow}
\usepackage[normalem]{ulem}
\usepackage{graphicx}
\usepackage{amsmath,amssymb}
\usepackage{dcolumn}
\usepackage{bm}
\usepackage{color}
\usepackage{appendix}
\newcommand{\f}{\frac}
\newcommand{\lt}{\left}

\newcommand{\n}{\nonumber}
\newcommand{\p}{\partial}
\newcommand{\rt}{\right}
\newcommand{\dd}{{\rm d}}

\newcommand{\arxgr}[1]{\href{http://arxiv.org/abs/#1}{{\ttfamily arXiv:#1[gr-qc]}}}
\newcommand{\arxph}[1]{\href{http://arxiv.org/abs/#1}{{\ttfamily arXiv:#1[hep-ph]}}}
\newcommand{\arxth}[1]{\href{http://arxiv.org/abs/#1}{{\ttfamily arXiv:#1[hep-th]}}}

\newcommand{\arxas}[1]{\href{http://arxiv.org/abs/#1}{{\ttfamily arXiv:#1[astro-ph.CO]}}}
\newcommand{\arxashe}[1]{\href{http://arxiv.org/abs/#1}{{\ttfamily arXiv:#1[astro-ph.HE]}}}
\newcommand{\arxasim}[1]{\href{http://arxiv.org/abs/#1}{{\ttfamily arXiv:#1[astro-ph.IM]}}}
\newcommand{\arxasga}[1]{\href{http://arxiv.org/abs/#1}{{\ttfamily arXiv:#1[astro-ph.GA]}}}
\newcommand{\Arxgr}[1]{\href{http://arxiv.org/abs/gr-qc/#1}{{\ttfamily arXiv:#1[gr-qc]}}}
\newcommand{\Arxth}[1]{\href{http://arxiv.org/abs/hep-th/#1}{{\ttfamily arXiv:#1[hep-th]}}}
\newcommand{\Arxas}[1]{\href{http://arxiv.org/abs/astro-ph/#1}{{\ttfamily arXiv:#1[astro-ph]}}}
\newcommand{\Arxph}[1]{\href{http://arxiv.org/abs/hep-ph/#1}{{\ttfamily arXiv:#1[hep-ph]}}}

\title{\boldmath Constraints on the inflaton potential from scalar-induced gravitational waves and primordial black holes}
\author[a]{Shi-Jie Wang,} 
\author[a,1]{Nan Li \note{Corresponding author.}}
\affiliation[a]{Department of Physics, College of Sciences, Northeastern University, Shenyang, 110819, China}
\emailAdd{2300203@stu.neu.edu.cn}
\emailAdd{linan@mail.neu.edu.cn}

\abstract{A plateau on the background inflaton potential $V_{\rm b}(\phi)$ can lead cosmic inflation into an ultraslow-roll phase, greatly enhancing the primordial power spectrum on small scales, and resulting in intensive scalar-induced gravitational waves (GWs) and abundant primordial black holes (PBHs). In this work, we construct an anti-symmetric perturbation $\delta V(\phi)$ on $V_{\rm b}(\phi)$ with three model parameters, the position, width, and slope of $\delta V(\phi)$, and constrain these parameters from the potential stochastic gravitational wave background (SGWB) in the NANOGrav 15-year data set. The GW spectrum from the supermassive black hole binaries (SMBHBs) with two model parameters, the amplitude and spectral index, is also investigated for comparison. We perform the Bayesian analysis in three steps with increasing number of model parameters, and obtain the allowed parameter ranges. When the constraints on PBH abundance are taken into account, these ranges become further narrower. We find that the increase of model parameters cannot significantly improve the Bayes factors, and the model with an almost perfect plateau on $V_{\rm b}(\phi)$ is favored. Moreover, the interpretation of the SGWB only via the GWs from SMBHBs is not preferred by the data. Two different forms of $V_{\rm b}(\phi)$ are considered, and the analogous results confirm the generality and robustness of our work.}

\keywords{Scalar-induced gravitational waves, Primordial black holes, Ultraslow-roll inflation, Bayesian analysis}

\begin{document}
\maketitle
\flushbottom

\section{Introduction} \label{sec:intro}

The detections of the gravitational waves (GWs) from compact binary mergers (e.g., the black hole binaries (BHBs) \cite{LIGOScientific:2016aoc, KAGRA:2021vkt} and the neutron star binaries \cite{LIGOScientific:2017vwq, LIGOScientific:2017ync}) have marked the beginning of an unprecedented multi-messenger epoch in modern cosmology. Moreover, in the early Universe, primordial GWs may also have different origins, such as cosmological phase transitions \cite{Kosowsky:1992vn, Grojean:2006bp, Caprini:2015zlo, Hindmarsh:2020hop}, topological defects \cite{Martin:1996ea, Figueroa:2012kw, Saikawa:2017hiv, Gouttenoire:2023ftk}, preheating \cite{Figueroa:2017vfa, Adshead:2018doq}, kination-domination \cite{Giovannini:1998bp, Figueroa:2019paj, Co:2021lkc, Gouttenoire:2021jhk}, and particle production \cite{Guzzetti:2016mkm, Peloso:2016gqs}. In addition, GWs can also be induced by scalar perturbations during cosmic inflation \cite{Baumann:2007zm, Bugaev:2009zh, Espinosa:2018eve, Kohri:2018awv, Yuan:2021qgz}. Altogether, these diverse sources collectively contribute to the stochastic GW background (SGWB), offering a new tool for probing the Universe and providing crucial information and insight for understanding astrophysics and cosmology beyond traditional electromagnetic methods \cite{Maggiore:1999vm, Caprini:2018mtu}. Therefore, the observation of the SGWB is the primary goal of various GW detectors that are designed for these faint but informative signals from a wide range of cosmic events. 

The recent observations from the European Pulsar Timing Array (EPTA) \cite{EPTA:2023fyk, EPTA:2023sfo}, Parkes PTA \cite{Reardon:2023gzh, Zic:2023gta}, Chinese PTA \cite{Xu:2023wog}, Meerkat PTA \cite{Miles:2024seg}, and especially North American Nanohertz Observatory for Gravitational Waves (NANOGrav) \cite{NANOGrav:2023gor, NANOGrav:2023hde, NANOGrav:2023hvm, NANOGrav:2023hfp}, have provided multiple pieces of evidence for an excess of the red common-spectrum signals at low frequencies (the nHz range) with respect to the Hellings--Downs correlations \cite{Hellings:1983fr}, supporting the potential existence of the isotropic SGWB. These signals follow a power-law spectrum, consistent with the GW spectrum from the orbital evolution of supermassive BHBs (SMBHBs) \cite{NANOGrav:2023hfp}. However, the spectral index deviates from the SMBHB-only model, suggesting environmental effects or other possible physical origins (e.g., cosmic (super)strings, first-order phase transitions, domain walls, large primordial fluctuations, and audible axions) \cite{NANOGrav:2023hvm}. These observations have aroused intensive research interests and debates in recent years \cite{Madge:2023dxc, Zu:2023olm, Vagnozzi:2023lwo, Franciolini:2023wjm, Franciolini:2023pbf, Wang:2023len, Addazi:2023jvg, Athron:2023mer, Jiang:2023qbm, Depta:2023uhy, Borah:2023sbc, Bi:2023tib, Niu:2023bsr, Ben-Dayan:2023lwd, Abe:2023yrw, Ghosh:2023aum, Bian:2023dnv, Servant:2023mwt, Wu:2023hsa, Choudhury:2023kam, Gouttenoire:2023bqy, Babichev:2023pbf, Zhang:2023nrs, Das:2023nmm, Liu:2023pau, Jiang:2023gfe, An:2023jxf, Frosina:2023nxu, He:2023ado, Cyr:2023pgw, Gangopadhyay:2023qjr, Liu:2023hpw, Sato-Polito:2023gym, Maity:2024odg, Chen:2024twp, Kume:2024adn, Sah:2024oyg, Smarra:2024kvv, Petrov:2024hec, Allen:2024mtn, Depta:2024ykq, Truant:2024aci, Liepold:2024woa, Qiu:2024sdd, Konstandin:2024fyo, Bernardo:2024bdc, Chen:2024hqc, Athron:2024fcj, Conzinu:2024cwl}. For instance, a multi-model analysis was performed on the above-mentioned models, with and without the SMBHB background, indicating that these models are capable of fitting the NANOGrav data as well as SMBHBs alone \cite{Ellis:2023oxs}. Also, in Ref. \cite{Figueroa:2023zhu}, the authors combined the NANOGrav and EPTA-DR2new data sets, and found that the audible axion or string signals fit the data better than SMBHBs. These studies pointed out alternative interpretations of the PTA data beyond the SMBHB-only model, and this is also the topic of our present work. 

In this work, we attribute the SGWB to the scalar-induced GWs (SIGWs), which are the second-order effect from the first-order scalar perturbations generated during inflation. If these scalar perturbations are large enough on small scales, they can also produce a significant number of primordial black holes (PBHs) at the same time \cite{hkkk, Carr:1974nx, Carr:1975qj}. These PBHs can explain a portion of the BHB merger events \cite{Khlopov:2008qy, Bird:2016dcv, Clesse:2016vqa, Sasaki:2016jop}, serve as the seeds of supermassive black holes at galactic centers \cite{Clesse:2015wea, Bernal:2017nec, Carr:2018rid, Yuan:2023bvh, Su:2023jno}, and more importantly, are a natural candidate of dark matter (DM) \cite{Carr:2016drx, Carr:2019kxo, Carr:2020gox, Carr:2020xqk, Green:2020jor, Escriva:2022duf}. For more recent researches on SIGWs and PBHs, especially in the light of the NANOGrav 15-year data set, see Refs. \cite{Li:2023qua, Huang:2023chx, Cai:2023dls, Inomata:2023zup, Gouttenoire:2023nzr, Unal:2023srk, Yi:2023mbm, Wang:2023sij, You:2023rmn, HosseiniMansoori:2023mqh, Cheung:2023ihl, Balaji:2023ehk, Yi:2023tdk, Bhaumik:2023wmw, Yi:2023npi, Harigaya:2023pmw, Li:2023xtl, Tagliazucchi:2023dai, Heydari:2023rmq, Nassiri-Rad:2023asg, Chang:2023vjk, Chen:2024fir, Ferreira:2024eru, Ianniccari:2024bkh, Domenech:2024rks, Perna:2024ehx, Papanikolaou:2024fzf, Escriva:2024ivo, Domenech:2024drm, Iovino:2024tyg, Afzal:2024hwj, Zhou:2024ncc, Datta:2024bqp, Kohri:2024qpd, Zhou:2024doz, Domenech:2024wao, Maji:2024cwv, Choudhury:2024ezx, Zhou:2024yke, Picard:2024ekd, Iovino:2024sgs, Allegrini:2024ooy, Wu:2024qdb, Chen:2025wcw, Wang:2025aon, Nandi:2025ihs} and the reviews \cite{Barack:2018yly, Ragavendra:2023ret, Carr:2023tpt, LISACosmologyWorkingGroup:2023njw}, with the references therein.

Usually, the single-field slow-roll (SR) inflation models lead to a nearly scale-invariant power spectrum of scalar perturbations, confirmed by the measurements of the cosmic microwave background (CMB) anisotropies on large scales \cite{Planck:2018jri}. However, the SIGWs from such a power spectrum are too weak to account for the SGWB. To have sufficiently large SIGWs, it is necessary to break the SR conditions by imposing a period of ultraslow-roll (USR) phase in cosmic inflation, during which the inflaton potential is rather flat, and the inflaton rolls down extremely slowly. In the USR stage, the SR parameters change dramatically, and the power spectrum on small scales is significantly amplified, generating intensive SIGWs and abundant PBHs simultaneously. Nevertheless, the position and duration of the USR phase must be carefully adjusted, so that it only influences the physics on small scales, without spoiling the large-scale CMB constraints. There are many mechanisms for the USR inflation, which can be realized by a inflection point or a saddle point on the inflaton potential \cite{Germani:2017bcs, Dimopoulos:2017ged, Di:2017ndc, Ashoorioon:2020hln}, in modified gravity theories \cite{Pi:2017gih, Lin:2020goi, Saburov:2024und}, or in supergravity models \cite{Wu:2021zta, Ishikawa:2021xya, Ishikawa:2024xjp}, etc. For a comprehension of the USR inflation and its effects on SIGWs and PBHs, see Refs. \cite{Ezquiaga:2017fvi, Cicoli:2018asa, Dalianis:2018frf, Cheng:2018qof, Mishra:2019pzq, Ragavendra:2020sop, Kefala:2020xsx, Wu:2021mwy, Ozsoy:2021pws, Hooshangi:2022lao, Franciolini:2022pav, Gu:2022pbo, Mu:2022dku, Mu:2023wdt, Ragavendra:2024yfp, Kristiano:2024vst, Autieri:2024wye, Aragam:2024nej, Tripathy:2024ngu, Ruiz:2024weh, Cielo:2024poz, Caravano:2024moy, Sharma:2024fbr, Fang:2025vhi}. In this paper, following our previous works in Refs. \cite{Wang:2021kbh, Liu:2021qky, Zhao:2023xnh, Zhao:2023zbg, Zhao:2024yzg, Su:2025mam}, we consider an anti-symmetric perturbation $\delta V(\phi)$ on the background inflaton potential to realize the USR phase. 

Our goal in this work is to determine the position, width, and slope of $\delta V(\phi)$ by the constraints from the SIGW spectrum and the PBH abundance. First, we adopt the NANOGrav 15-year data set, utilize a modified version of the \texttt{SIGWfast} code \cite{Witkowski:2022mtg} to compute the SIGW spectrum, and perform Bayesian analysis \cite{Thrane:2018qnx} by the \texttt{PTArcade} code \cite{Mitridate:2023oar}. Second, the model parameters extracted from the SIGW spectrum must lead to reasonable PBH abundance in relevant mass range at the same time, and this demand will further shrink the allowed parameter space. Third, as there is also possibility that the SGWB is generated from SMBHBs \cite{NANOGrav:2023gor, NANOGrav:2023hvm}, we still need to consider the scenario that the SGWB origins from both SIGWs and SMBHBs. These three points will constitute the main body of our work. More importantly, we do not assume any simple form of the primordial power spectrum as done in many previous literature, but constrain the model parameters of the inflaton potential directly from the GW data. This operation is also supported by Ref. \cite{LISACosmologyWorkingGroup:2025vdz}, where the authors studied different methods (template based, agnostic, and ab initio) to constrain the USR phase, and found that using the explicit USR construction leads to the minimum residuals.

This paper is organized as follows. In Sec. \ref{sec:USR}, we introduce two different background inflaton potentials, show the power spectrum of the primordial curvature perturbation, and calculate the PBH abundance. In Sec. \ref{sec:SIGWs}, we derive the SIGW spectrum. The NANOGrav 15-year data set and the relevant Bayesian analysis method are reviewed in Sec. \ref{sec:bayes}. With all these preparations, we constrain our inflation models in Sec. \ref{sec:Constraints} and App. \ref{sec:app}, with two model parameters, three model parameters, and SMBHBs taken into account, respectively. The Bayes factors between different models are also calculated and discussed in detail. We conclude in Sec. \ref{sec:Conclusion}. We work in the natural system of units and set $c=\hbar=k_{\mathrm{B}}=1$.

\section{USR inflation and PBHs} \label{sec:USR}

In this section, we discuss the equation of motion for the primordial curvature perturbation $\mathcal{R}$, introduce its power spectrum ${\cal P}_{\mathcal{R}}(k)$, and show the calculation of PBH abundance $f_{\rm PBH}$.

\subsection{Basic equations} \label{sec:USR:Basic}

We start from the action of the single-field inflation,
\begin{align}
S=\int\mathrm{d}^4x\,\sqrt{-g}\lt[\f{m_{\text{P}}^2}{2}R-\f{1}{2}\p_\mu\phi\p^\mu\phi-V(\phi)\rt], \n
\end{align}
where $m_{\rm P}=1/\sqrt{8\pi G}=2.435\times 10^{18}$ GeV is the reduced Planck mass, $R$ is the Ricci scalar, $\phi$ is the inflaton field, and $V(\phi)$ is its potential, respectively. By variation of $S$, the equation of motion for $\phi$ can be obtained as the Klein--Gordon equation,
\begin{align}
\ddot{\phi}+3H\dot{\phi}+V_{,\phi}=0, \label{KGeq}
\end{align}
where $H$ is Hubble expansion rate. 

During inflation, it is more convenient to use the number of $e$-folds $N$ to describe cosmic expansion, defined as $\mathrm{d}N=H(t)\,\mathrm{d}t=\mathrm{d}\ln{a}(t)$, where $t$ is the cosmic time, $a(t)$ is the scale factor, and $H=\dot{a}/a$. In addition, to characterize the SR behaviors of the inflaton, two SR parameters can be introduced as 
\begin{align}
\varepsilon&=-\f{\dot{H}}{H^{2}}=\f{\phi_{,N}^{2}}{2m_{\mathrm{P}}^{2}},\quad \eta=-\f{\ddot{\phi}}{H\dot{\phi}}=\f{\phi_{,N}^{2}}{2m_{\mathrm{P}}^{2}}-\f{\phi_{,NN}}{\phi_{,N}}. \label{SRP}
\end{align}
With these parameters, Eq. (\ref{KGeq}) can be reexpressed as 
\begin{align}
\phi_{,NN}+(3-\varepsilon)\phi_{,N}+\f{1}{H^2}V_{,\phi}=0, \n
\end{align}
and the Friedmann equation for cosmic expansion can be written as 
\begin{align}
H^2=\f{V}{(3-\varepsilon)m_{\rm P}^2}. \n
\end{align}

Then, we move on to the perturbations on the background spacetime. In Newtonian gauge, the perturbed metric reads
\begin{align}
\mathrm{d}s^{2}&=-(1+2\Psi)\,\mathrm{d}t^{2}+a^{2}(t)\lt[(1-2\Psi)\delta_{ij}+\f{1}{2}h_{ij}\rt]\,\mathrm{d}x^{i}\mathrm{d}x^{j}, \label{perturbed metric}
\end{align}
where $\Psi$ is the first-order scalar perturbation, $h_{ij}$ is the second-order tensor perturbation, and we have neglected the vector perturbation and anisotropic stress here. A more frequently used gauge-invariant scalar perturbation is the primordial curvature perturbation $\mathcal{R}$, 
\begin{align}
\mathcal{R}=\Psi+\f{H}{\dot{\phi}}\,\delta\phi, \n
\end{align}
where $\delta\phi$ is the perturbation of the inflaton field. The equation of motion for its Fourier mode $\mathcal{R}_k$ is the Mukhanov--Sasaki equation \cite{Sasaki:1983kd, Mukhanov},
\begin{align}
\mathcal{R}_{k,NN}+(3+\varepsilon-2\eta)\mathcal{R}_{k,N}+\f{k^2}{H^2e^{2N}}\mathcal{R}_k=0. \label{MS}
\end{align}

\subsection{Inflaton potentials} \label{sec:USR:Power spectrum}

The Fourier mode $\mathcal{R}_k$ can be obtained by numerically solving Eqs. \eqref{KGeq}, \eqref{SRP}, and \eqref{MS}, which requires us to specify the form of the inflaton potential $V(\phi)$. Following our previous works in Refs. \cite{Wang:2021kbh, Liu:2021qky, Zhao:2023xnh, Zhao:2023zbg, Zhao:2024yzg, Su:2025mam}, we decompose $V(\phi)$ as the sum of its background potential $V_{\rm b}(\phi)$ and a perturbation $\delta V(\phi)$ on it,
\begin{align}
V(\phi) = V_\mathrm{b}(\phi) + \delta V(\phi). \n
\end{align}

For the generality of our discussions, we explore two different background potentials in this work. The first is the Kachru--Kallosh--Linde--Trivedi (KKLT) potential \cite{Kachru:2003aw},
\begin{align}
V_{\rm b} (\phi)=V_0\f{\phi^2}{\phi^2+(m_\mathrm{P}/2)^2}, \n 
\end{align}
where the parameter $V_0$ indicates the energy scale of inflation. The second is the Starobinsky potential \cite{Starobinsky}, which origins from an $R^2$-modification in the Einstein--Hilbert action and can be mapped to a scalar field $\phi$ minimally coupled to gravity via a conformal transformation,
\begin{align}
V_{\rm b}(\phi)=V_0\left(1-e^{\sqrt{2/3}\phi/m_{\rm P}}\right)^2. \n
\end{align}
Here, we set $V_0$ the same as that in the KKLT model, so that the inflatons roll down the potentials from the same energy scale.

\begin{figure}[htb]
\includegraphics[width=0.495\textwidth]{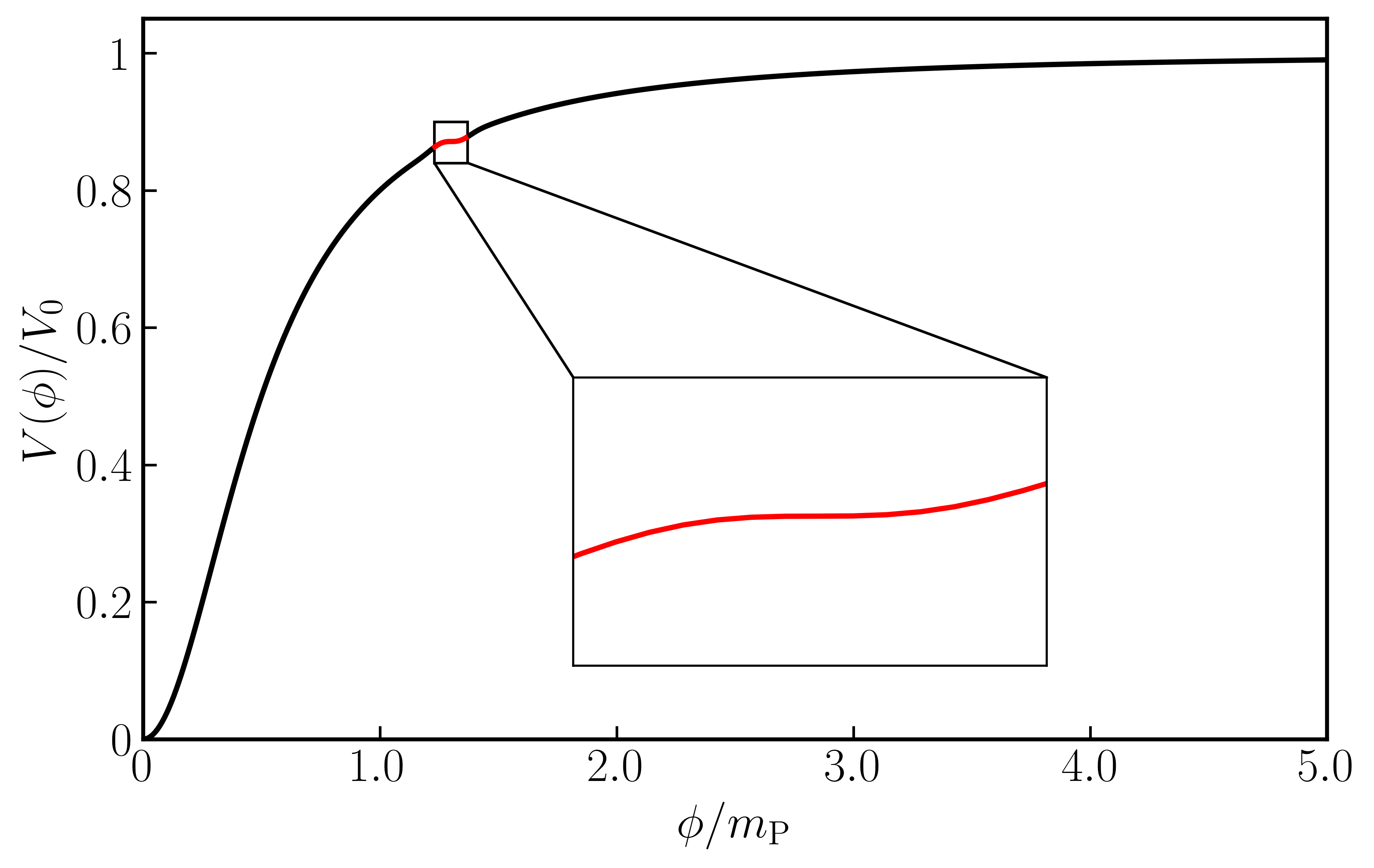}
\includegraphics[width=0.495\textwidth]{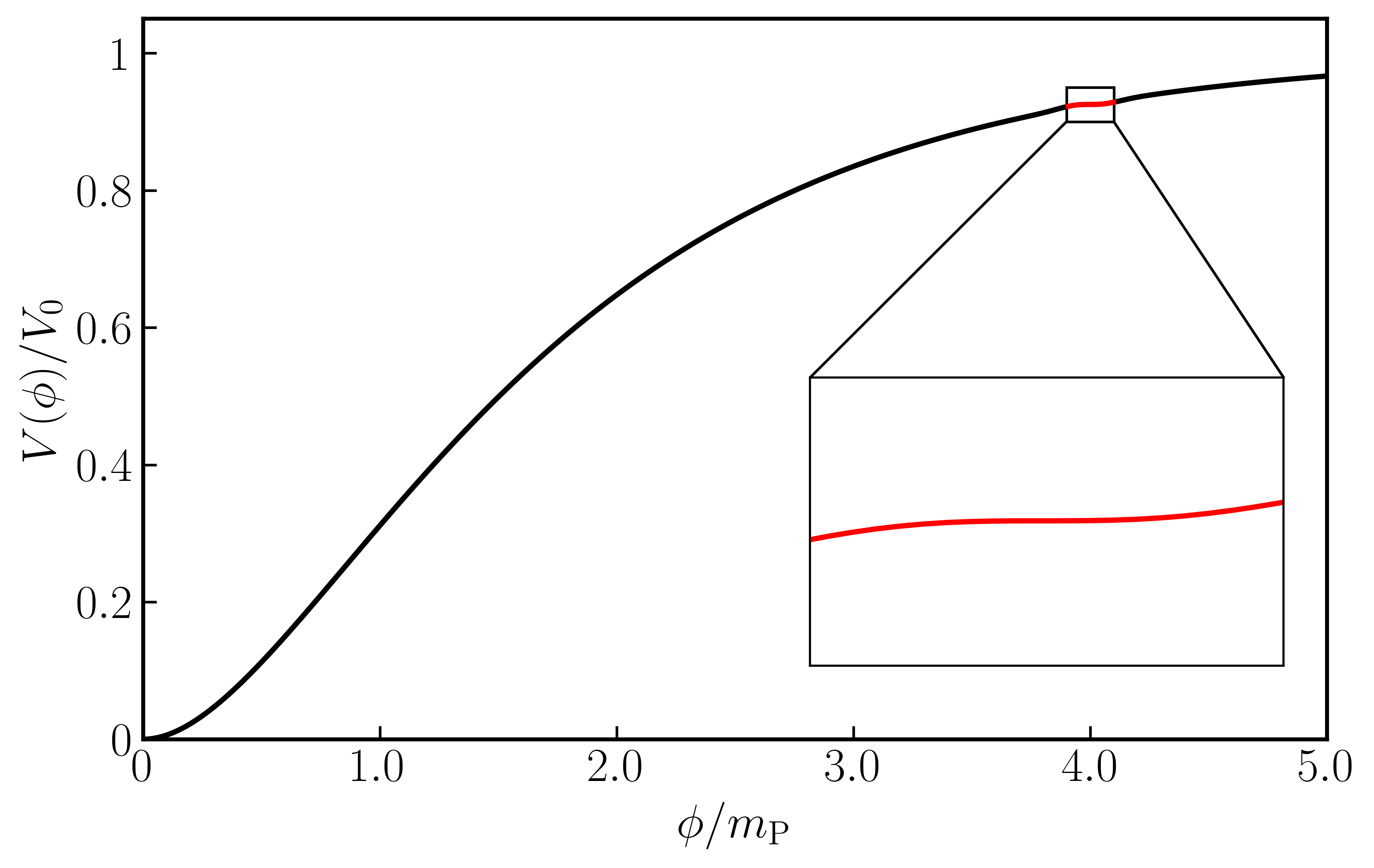}
\caption{The inflaton potentials for the KKLT model (the left panel) and the Starobinsky model (the right panel), with the relevant USR regions shown in the insets. The shapes of these two potentials are similar, while the Starobinsky potential decreases relatively slowly. The inflatons roll down these two potentials from the same energy scale at $V_0/m^4_{\text{P}}=10^{-10}$ and pass through the USR regions at different positions, resulting in abundant PBHs and intensive SIGWs.}
\label{fig:potential}
\end{figure}

Then, we parameterize the perturbation $\delta V(\phi)$ in an anti-symmetric form as
\begin{align}
\delta V(\phi)=-A(\phi-\phi_0)\exp\lt[-\f{(\phi-\phi_0)^2}{2\sigma^2}\rt], \label{VVV}
\end{align}
where $A$, $\phi_0$, and $\sigma$ are three model parameters that determine the amplitude, position, and width of $\delta V(\phi)$, respectively. For convenience, we reexpress $A$ as
\begin{align}
A=V_{\text{b},\phi}(\phi_0)(1+A_0), \label{VAA}
\end{align}
with $V_{\rm b,\phi}(\phi_0)=\dd V_{\rm b}(\phi)/\dd\phi|_{\phi_0}$ denoting the slope of the background potential $V_{\rm b}(\phi)$ at the position $\phi_0$. By this means, $A_0$ characterizes the deviation of $V(\phi)$ from a perfect plateau at $\phi_0$. We plot the two inflaton potentials in Fig. \ref{fig:potential}, with the USR regions shown in the insets.

Furthermore, in the KKLT model, we set $V_0/m^4_{\text{P}}=10^{-10}$, $\phi_{\rm i}/m_\text{P}=3.30$, and $\phi_{{\rm i},N}/m_\text{P}=-0.0137$ as the initial conditions for inflation, such that there can be a nearly scale-invariant power spectrum, with the spectral index $n_{\rm s}=0.9600$ and a relatively small tensor-to-scalar ratio $r=0.00321$, favored by the CMB observations on large scales \cite{Planck:2018jri}. Similarly, in the Starobinsky model, the corresponding initial conditions read $V_0/m^4_{\text{P}}=10^{-10}$, $\phi_{\rm i}/m_\text{P}=5.80$, and $\phi_{{\rm i},N}/m_\text{P}=-0.0137$, and we have $n_{\rm s}=0.9669$ and $r=0.00321$, also consistent with the CMB constraints. 



\subsection{Power spectra and PBH abundance} \label{sec:USR:PBH abundance}

To obtain the PBH abundance, we introduce the dimensionless power spectrum $\mathcal{P}{_{\mathcal{R}}}(k)$ of ${\cal R}_k$,
\begin{align}
\mathcal{P}_{\mathcal{R}}(k)=\f{k^3}{2\pi^2}|\mathcal{R}_k|^2\bigg|_{k\ll aH}. \label{Pspec}
\end{align}
Here, we should emphasize that, as ${\cal R}_k$ can still evolve after exiting the Hubble horizon during the USR inflation, $\mathcal{P}_{\mathcal{R}}(k)$ must be evaluated at the end of inflation, with the relevant scale satisfying $k\ll aH$. 

Furthermore, we need to link $\mathcal{P}_{\mathcal{R}}(k)$ to the dimensionless power spectrum $\mathcal{P}_{\delta}(k)$ of the primordial density contrast $\delta$. In the radiation-dominated (RD) era, at the linear order, we have \cite{Green:2004wb}
\begin{align}
\mathcal{P}_{\delta}(k)=\frac{16}{81}\left(\frac{k}{aH}\right)^{4}\mathcal{P}_{\mathcal{R}}(k). \n
\end{align}
By this relation, the $i$-th spectral moment $\sigma_i$ of the density contrast $\delta$ can be obtained as
\begin{align}
\sigma_{i}^{2}(R)&=\int_{0}^{\infty}\f{\mathrm{d}k}{k}\, k^{2i}\widetilde{W}^{2}(k,R)\mathcal{P}_{\delta}(k) 
=\f{16}{81}\int_{0}^{\infty}\f{\mathrm{d}k}{k}\, k^{2i}\widetilde{W}^{2}(k,R)(kR)^{4}\mathcal{P}_{\mathcal R}(k), \label{spectral moment}
\end{align}
where $i = 0,1,2,\cdots$, and $\widetilde{W}(k,R)$ is the window function in the Fourier space, which smooths the perturbation over some physical scale $R$, often chosen as $R=1/(aH)$. In this work, we choose Gaussian window function as $\widetilde{W}^{2}(k,R)=e^{-k^2R^2/2}$, which prevents the issues related to the non-differentiability and divergence of the integral in the large-$k$ limit in Eq. \eqref{spectral moment}.

For PBH formation, in the Carr--Hawking collapse model \cite{Carr:1974nx}, the PBH mass $M$ is related to the horizon mass $M_{\mathrm{H}}$ as
\begin{align}
M=\kappa M_{\mathrm{H}}=\kappa\f{4\pi m^2_{\text{P}}}{H}, \n
\end{align}
where $\kappa$ is the collapse efficiency. Then, considering the conservation of entropy in the adiabatic cosmic expansion, we can obtain \cite{Carr:2020gox}
\begin{align}
\f{M}{M_\odot}=1.13\times10^{15}\lt(\f{\kappa}{0.2}\rt)\lt(\f{g_*}{106.75}\rt)^{-1/6}\lt(\f{k_*}{k_{\mathrm{PBH}}}\rt)^2, \label{M_k}
\end{align}
where $M_{\odot}=1.989\times10^{30}$ kg is the solar mass, $g_*$ is the effective number of relativistic degrees of freedom of energy density, $k_*=0.05~\text{Mpc}^{-1}$ is the CMB pivot scale for the Planck satellite experiment \cite{Planck:2018jri}, and $k_\text{PBH}=1/R$ is the wave number of the PBH that exits the Hubble horizon. Below, we set $\kappa = 0.2$ and $g_* = 106.75$ in the RD era \cite{Carr:1975qj}. From Eq. \eqref{M_k}, all spectral moments $\sigma_i(R)$ can be reexpressed in terms of the PBH mass as $\sigma_i(M)$.

At the epoch of PBH formation in the RD era, its mass fraction $\beta_{\rm{PBH}}(M)$ is defined as
\begin{align}
\lt.\beta_{\mathrm{PBH}}(M)=\f{\rho_{\mathrm{PBH}}(M)}{\rho_{\mathrm{R}}}\rt|_{\rm formation}, \n 
\end{align}
where $\rho_{\mathrm{PBH}}(M)$ and $\rho_{\mathrm{R}}$ are the energy densities of PBH and radiation. Then, at the present time, the PBH abundance $f_{\rm PBH}(M)$ is defined as
\begin{align}
f_{\mathrm{PBH}}(M)=\lt.\f{\rho_{\mathrm{PBH}}(M)}{\rho_{\mathrm{DM}}}\rt|_{\rm today}, \n 
\end{align}
where $\rho_{\mathrm{DM}}$ is the energy density of DM. If we focus on monochromatic PBH mass, $f_{\mathrm{PBH}}(M)$ is naturally proportional to $\beta_{\mathrm{PBH}}(M)$ \cite{Carr:2020gox}, 
\begin{align}
f_{\mathrm{PBH}}(M)=1.68\times10^8\lt(\f{M}{M_\odot}\rt)^{-1/2}\beta_{\mathrm{PBH}}(M). \n
\end{align}

The PBH mass fraction $\beta_{\mathrm{PBH}}(M)$ can be calculated in different ways, corresponding to taking the spectral moments $\sigma_i$ at different orders. In the general peak theory \cite{Bardeen:1985tr}, all $\sigma_0$, $\sigma_1$, and $\sigma_2$ are needed, but the relevant calculations are rather tedious. A very convenient and effective simplification is the Green--Liddle--Malik--Sasaki approximation \cite{Green:2004wb}, which only consults $\sigma_0$ and $\sigma_1$ but yields almost the same results as peak theory. Hence, in this work, we will follow this approximation, which gives
\begin{align}
\beta_\text{PBH}(M)=\f{1}{\sqrt{2\pi}}Q^3(\nu_{\rm c}^2-1)e^{-\nu_{\rm c}^2/2}, \n
\end{align}
where $Q=R\sigma_1/(\sqrt{3}\sigma_0)$, $\nu_{\text{c}}=\delta_{\text{c}}/\sigma_{0}$ is the relative density contrast, and $\delta_{\rm c}$ is the threshold of $\delta$ for PBH formation, which depends on many physical ingredients, such as the shape of the collapsing region \cite{Musco:2018rwt, Musco:2004ak, Harada:2013epa, Nakama:2013ica, Escriva:2019nsa, Musco:2020jjb}, and the equation of state and the anisotropy of the cosmic medium \cite{Musco:2021sva, Papanikolaou:2022cvo, Stamou:2023vxu}. In this work, we follow Ref. \cite{Harada:2013epa} and choose $\delta_{\rm c}=0.414$.

\section{SIGWs} \label{sec:SIGWs}

In this section, we provide a concise review of the derivation of the SIGW spectra in the RD era and at the present time.

\subsection{Basic equations} \label{sec:SIGWs:Basic}

The equation of motion for the second-order tensor perturbation $h_{ij}$ in Eq. \eqref{perturbed metric} reads \cite{Baumann:2007zm}
\begin{align}
h_{ij}^{\prime\prime}+2\mathcal{H}h_{ij}^{\prime}-\nabla^2h_{ij}=-4{\mathcal{T}}_{ij}^{lm}\mathcal{S}_{lm}, \label{Einstein eq}
\end{align}
where ${\mathcal{T}}_{ij}^{lm}$ is the projection operator, $\mathcal{S}_{lm}$ is the source term, a prime denotes the derivative with respect to the conformal time $\tau$ defined as $\mathrm{d}\tau=\mathrm{d}t/a(t)$, and ${\cal H}=a'/a$. In the RD era, using the conformal Newtonian gauge, we have 
\begin{align}
\mathcal{S}_{ij} &= 4\Psi\p_i\p_j\Psi +2\p_i\Psi\p_j\Psi-\f{1}{\mathcal{H}^2}\p_i(\Psi^{\prime}+\mathcal{H}\Psi)\p_j(\Psi^{\prime}+\mathcal{H}\Psi). \n 
\end{align}

In general, the Fourier transform of $h_{ij}$ is
\begin{align}
h_{ij}(\tau, \mathbf{x})=\int \f{\mathrm{d}^3k}{(2\pi)^{3/2}} \, e^{i\mathbf{k} \cdot \mathbf{x}}\lt[h_{\mathbf{k}}^{+}(\tau) {e}_{ij}^{+}(\mathbf{k})+h_{\mathbf{k}}^{\times}(\tau){e}_{i j}^{\times}(\mathbf{k})\rt], \n
\end{align}
where $h_{\mathbf{k}}^{+}(\tau)$ and $h_{\mathbf{k}}^{\times}(\tau)$ are the Fourier components of $h_{ij}(\tau,{\bf x})$, and ${e}_{ij}^{+}(\mathbf{k})$ and ${e}_{ij}^{\times}(\mathbf{k})$ are two time-independent polarization tensors that can be expressed in terms of two orthogonal basis vectors $\mathbf{e}$ and $\bar{\mathbf{e}}$ as
\begin{align}
{e}_{i j}^{+}(\mathbf{k}) &= \f{1}{\sqrt{2}}\lt[{e}_{i}(\mathbf{k}) {e}_{j}(\mathbf{k})-\overline{{e}}_{i}(\mathbf{k}) \overline{{e}}_{j}(\mathbf{k})\rt],\quad {e}_{i j}^{\times}(\mathbf{k}) = \f{1}{\sqrt{2}}\lt[{e}_{i}(\mathbf{k}) \overline{{e}}_{j}(\mathbf{k})+\overline{{e}}_{i}(\mathbf{k}) {e}_{j}(\mathbf{k})\rt]. \n
\end{align}

In the Fourier space, Eq. \eqref{Einstein eq} becomes
\begin{align}
h''_{\mathbf{k}}+2\mathcal{H}h^{\prime}_{\mathbf{k}}+k^2h_{\mathbf{k}}=\mathcal{S}(\tau, \mathbf{k}), \label{Einstein feq}
\end{align}
where $\mathcal{S}(\tau,\mathbf{k})$ is the Fourier transform of the source term $\mathcal{S}_{lm}(\tau, \mathbf{x})$, and we have ignored the polarization indices $+$ and $\times$. Furthermore, we decompose the Fourier mode of the scalar perturbation $\Psi$ as $\Psi_{\mathbf{k}}(\tau)=\Psi(k\tau)\psi_{\mathbf{k}}$, where $\psi_{\mathbf{k}}$ is the initial value, and $\Psi(k\tau)$ is the normalized transfer function. In the RD era, we have \cite{Kohri:2018awv}
\begin{align}
\Psi(k\tau)=\f{9}{(k\tau)^2}\lt[\f{\sin(k\tau/\sqrt{3})}{k\tau/\sqrt{3}}-\cos(k\tau/\sqrt{3})\rt]. \n
\end{align}
In this way, the source term can be reexpressed as
\begin{align}
{\cal S}(\tau,\mathbf{k})=\int\f{\mathrm{d}^{3}k}{(2\pi)^{3/2}} \,{e}^{ij}(\mathbf{k})\widetilde{k}_i\widetilde{k}_jf(\tau,\mathbf{k},\widetilde{\mathbf{k}})\psi_{\mathbf{k}}\psi_{\mathbf{k}-\widetilde{\mathbf{k}}}, \n
\end{align}
where the source function $f(\tau,\mathbf{k}, \widetilde{\mathbf{k}})$ is \cite{Domenech:2021ztg}
\begin{align}
f(\tau,{\bf k}, \widetilde{\bf k})&= 12\Psi(|\widetilde{\mathbf{k}}|\tau) \Psi(|\mathbf{k}-\widetilde{\bf k}|\tau)+4\tau^2\Psi^{\prime}(|\widetilde{\mathbf{k}}|\tau)\Psi^{\prime}(|\mathbf{k}-\widetilde{\mathbf{k}}|\tau) \n\\
&\quad +4\tau\lt[\Psi'(|\widetilde{\mathbf{k}}|\tau) \Psi(|\mathbf{k}-\widetilde{\mathbf{k}}|\tau) + \Psi(|\widetilde{\mathbf{k}}|\tau)\Psi'(|\mathbf{k}-\widetilde{\mathbf{k}}|\tau)\rt]. \n
\end{align}

With all above preparations, we can solve Eq. (\ref{Einstein feq}) by the Green function method,
\begin{align}
h_{\mathbf{k}}(\tau)=\f{1}{a(\tau)}\int \mathrm{d}\widetilde{\tau}\, G_{k}(\tau,\widetilde{\tau}) a(\widetilde{\tau})\mathcal{S}(\widetilde{\tau},\mathbf{k}), \n
\end{align}
where the Green function is $G_{k}(\tau,\widetilde{\tau})=\sin{[k(\tau-\widetilde{\tau})]/k}$.

\subsection{SIGW spectra} \label{sec:SIGWs:spectrum}

The SIGW spectrum $\Omega_{\mathrm{GW}}$ is defined as the GW energy density $\rho_{\mathrm{GW}}$ per logarithmic wave number as
\begin{align}
\Omega_\mathrm{GW}(\tau,k)=\f{1}{\rho_\mathrm{c}}\f{\mathrm{d}\rho_\mathrm{GW}(\tau,k)}{\mathrm{d}\ln k}, \n 
\end{align}
where $\rho_\mathrm{c}$ is the critical energy density of the Universe. In the transverse--traceless gauge, $\Omega_{\rm GW}(\tau,k)$ can be written as \cite{Isaacson:1968hbi, Isaacson:1968zza}
\begin{align}
\Omega_{\mathrm{GW}}(\tau,k)=\f{1}{24}\lt(\f{k}{\mathcal{H}}\rt)^2\overline{\mathcal{P}_h(\tau,k)}, \n 
\end{align}
where $\overline{(\cdots)}$ denotes the oscillation average, and $\mathcal{P}_h(\tau,k)$ is the dimensionless power spectrum of the tensor perturbation $h_\mathbf{k}$, defined as
\begin{align}
\mathcal{P}_h(\tau,k)=\f{k^3}{2\pi^2} \delta_{\rm D}(\mathbf{k}+\widetilde{\mathbf{k}})\langle h_\mathbf{k}(\tau)h_{\widetilde{\mathbf{k}}}(\tau)\rangle, \n 
\end{align}
where $\delta_{\rm D}$ is the Dirac $\delta$ function, and the two-point correlation function $\langle h_\mathbf{k}(\tau)h_{\widetilde{\mathbf{k}}}(\tau)\rangle$ reads
\begin{align}
\langle h_{\mathbf{k}}(\tau)h_{\widetilde{\mathbf{k}}}(\tau)\rangle&=\int\f{\mathrm{d}^{3}q\mathrm{d}^{3}\widetilde{q}}{(2\pi)^{3}}\,e^{ij}({\bf k})e^{lm}({\widetilde{\bf k}}) q_iq_j\widetilde{q}_l\widetilde{q}_mI(\tau,\mathbf{k},\mathbf{q}) I(\tau,\widetilde{\mathbf{k}},\widetilde{\mathbf{q}})\langle\psi_{\mathbf{q}}\psi_{\mathbf{k}-\mathbf{q}}\psi_{\widetilde{\mathbf{q}}}\psi_{\widetilde{\mathbf{k}}-\widetilde{\mathbf{q}}}\rangle, \label{4dian}
\end{align}
with $I(\tau,\mathbf{k},\mathbf{q})$ being the kernel function,
\begin{align}
I(\tau,\mathbf{k},\mathbf{q}) =\int\mathrm{d}\widetilde{\tau}\ \f{a(\widetilde{\tau})}{a(\tau)}G_k(\tau,\widetilde{\tau})f(\widetilde{\tau},\mathbf{k},\mathbf{q}). \n
\end{align}

Furthermore, by the Wick theorem, the four-point correlation function $\langle\psi_{\mathbf{q}}\psi_{\mathbf{k}-\mathbf{q}}\psi_{\widetilde{\mathbf{q}}}\psi_{\widetilde{\mathbf{k}}-\widetilde{\mathbf{q}}}\rangle$ in Eq. \eqref{4dian} can be decomposed as the sum of the products of two-point correlation functions. It is convenient to introduce three dimensionless variables as $x=k\tau$, $u=|\mathbf{k}-\widetilde{\mathbf{k}}|/k$, and $v=|\widetilde{\mathbf{k}}|/k$. By this means, $\mathcal{P}_h(\tau,k)$ can be obtained as 
\begin{align}
\mathcal{P}_{h}(\tau,k)=4\int_{0}^{\infty}\mathrm{d}v \int_{|1-v|}^{1+v}\mathrm{d}u \,\lt[\f{4v^{2}-(1+v^{2}-u^{2})^{2}}{4uv}\rt]^{2}\mathcal{I}^{2}(x,u,v)\mathcal{P}_{\mathcal{R}}(ku)\mathcal{P}_{\mathcal{R}}(kv), \n 
\end{align}
where $\mathcal{I}(x,u,v)= I(\tau,\mathbf{k},\widetilde{\mathbf{k}})k^2$ is the kernel function in terms of the dimensionless variables $x$, $u$, and $v$ \cite{Ananda:2006af}. In the RD era, the oscillation average of $\mathcal{I}^{2}(x,u,v)$ in the late-time limit of $x\rightarrow \infty$ is \cite{Kohri:2018awv}
\begin{align}
\overline{\mathcal{I}^{2}(x\rightarrow\infty,u,v)}&= \f{1}{2}\lt[\f{3(u^{2}+v^{2}-3)^{2}}{4u^{3}v^{3}x}\rt]^{2}\Bigg\{\pi^{2}(u^{2}+v^{2}-3)^{2} \Theta(u+v-\sqrt{3}) \n\\ 
&\quad +\lt[\lt(u^2+v^2-3\rt)\ln\lt|\f{3-(u+v)^2}{3-(u-v)^2}\rt|-4uv\rt]^2\Bigg\}, \n
\end{align}
where $\Theta$ is the Heaviside step function. We may further define two new variables as $s=u+v$ and $t=u-v$. Taking into account $\mathcal{H}=1/\tau$ in the RD era, we obtain the SIGW spectrum $\Omega_{\mathrm{GW}}(\tau,k)$ as
\begin{align}
\Omega_{\mathrm{GW}}(\tau,k) &=12\int_0^1\mathrm{d}t\int_1^\infty \mathrm{d}s\, \f{t(t^2-1)^2(s^2-1)^2(s^2+t^2-6)^4}{(s^2-t^2)^8} \n\\
&\quad\times\lt\{\lt[\ln\lt|\f{t^2-3}{s^2-3}\rt|+\f{2(s^2-t^2)}{s^2+t^2-6}\rt]^2+\pi^2\Theta(s-\sqrt{3})\rt\}\mathcal{P}_{\mathcal{R}} \lt(\f k2(s+t)\rt)\mathcal{P}_{\mathcal{R}}\lt(\f k2(s-t)\rt). \n
\end{align}
Moreover, there is a simple relation between the wave number $k$ and the frequency $f$ of the SIGW \cite{Xu:2019bdp},
\begin{align}
f\approx1.5\times10^{-15}\f{k}{\mathrm{Mpc}^{-1}}\,\mathrm{Hz}. \n
\end{align}
From this correspondence, we finally achieve the SIGW spectrum $\Omega_{\mathrm{GW}}(\tau,f)$
from $\Omega_{\mathrm{GW}}(\tau,k)$. 

Last, the SIGW spectrum at present time $\Omega_{\mathrm{GW}}(\tau_0,f)$ can be related to $\Omega_{\mathrm{GW}}(\tau_{\rm c},f)$ as
\begin{align}
\Omega_{\mathrm{GW}}(\tau_0,f)h^2=\Omega_{\rm R}^0h^2\f{g_{*}(f)}{g_{*}^{0}}\lt(\f{g_{*s}^{0}}{g_{*s}(f)}\rt)^{4/3}\Omega_{\mathrm{GW}}\lt(\tau_{\rm c},f\rt), \n
\end{align}
where $\Omega_{\mathrm{R}}^0$ is the present energy density fraction of radiation, $g_*$ and $g_{*s}$ are the effective relativistic degrees of freedom that contribute to radiation energy and entropy densities, and $\tau_{\rm c}$ is some time after $\Omega_{\rm{GW}}(\tau,f)$ has become constant, as SIGWs can be regarded as a portion of radiation, so $\Omega_{\rm{GW}}(\tau_{\rm c},f)$ is an asymptotic constant during the RD era. In this work, we take the values $\Omega_{\mathrm{R}}^0h^2 = 4.184\times10^{-5}$, $g_*^0=3.383$, and $g_{*s}^0=3.930$, respectively \cite{Saikawa:2020swg}. 

\section{Bayesian analysis} \label{sec:bayes}

In this section, we provide the elements of Bayesian analysis, including the likelihood function, prior and posterior distributions, and the Bayes factor between different models, respectively. The NANOGrav 15-year data set is also reviewed.

\subsection{Likelihood function} \label{sec:bayes:Likelihood}

The likelihood function $\mathcal{L}$ is related to the timing residual $\delta t$ of pulsars and can be expressed as \cite{Mitridate:2023oar, NANOGrav:2023gor, NANOGrav:2023hvm}
\begin{align}
\delta t = n + Fa + M \epsilon, \label{like}
\end{align}
where $\delta t$, $n$, $a$, and $\epsilon$ are column vectors, and $F$ and $M$ are matrices. Below, we explain Eq. (\ref{like}) in more detail. 

In Eq. (\ref{like}), $n$ is the white noise. The next term $Fa$ describes the time-correlated stochastic process, including the intrinsic pulsar red noise and the GW signals. First, $F$ is the Fourier basis matrix, composed of sines and cosines calculated from the time of arrival of the pulses with the frequency at $f_i=i/T_{\text{obs}}$ ($T_{\text{obs}}=16.03$ yr is the time of arrival extent). Second, $a$ is the amplitude, following a zero-mean normal distribution with the covariance matrix $\langle aa^{\rm T} \rangle=\phi$, which is given by
\begin{align}
[\phi]_{(a i)(b j)}=\delta_{ij}(\Gamma_{a b}\Phi_{i}+\delta_{ab}\varphi_{a,i}). \label{relation}
\end{align}
In Eq. (\ref{relation}), the subscripts $a$ and $b$ correspond to the pulsars, $i$ and $j$ represent the frequency harmonics, $\Gamma_{ab}$ is the GW background overlap reduction function, denoting the mean correlation between pulsars $a$ and $b$ based on their angular separation in the sky, and the model-dependent coefficients $\Phi_{i}$ can be related to the GW spectrum as
\begin{align}
\Omega_{\mathrm{GW}}(f)h^2=\f{h^2}{\rho_{\rm c}^0}\f{\dd \rho_{\mathrm{GW}}(f)} {\dd\ln f} =\f{8\pi^4f^5}{H_0^2/h^2} \f{\Phi(f)}{\Delta f}, \n 
\end{align}
where $H_0=100h~{\rm km}~{\rm s}^{-1}~{\rm Mpc}^{-1}$ is the Hubble constant, $\Phi(f)$ is defined as $\Phi_i=\Phi(i/T_{\text{obs}})$, and $\Delta f = 1/T_{\text{obs}}$ is the width of the frequency bin. Moreover, in Eq. (\ref{relation}), the spectral component $\varphi_{a,i} =\varphi_a(i/T_{\text{obs}})$ of the intrinsic pulsar noise is set in a power-law form as
\begin{align}
\varphi_a(f) = \f{A_a^2}{12\pi^2}\f{1}{T_{\text{obs}}}\lt(\f{f}{\text{yr}^{-1}}\rt)^{-\gamma_a}\,{\rm yr}^3, \n
\end{align}
where $A_a$ is the intrinsic noise amplitude, and $\gamma_a$ is the spectral index. Last, the term $M\epsilon$ in Eq. (\ref{like}) shows the uncertainty in the timing model, in which $M$ is the design-matrix basis, and $\epsilon$ is the coefficient. 

We can marginalize over $a$ and $\epsilon$ to obtain the likelihood function ${\cal L}$ that depends only on the parameters describing the red noise covariance matrix $\langle aa^{\rm T}\rangle$,
\begin{align}
\mathcal{L}(\delta t|\phi)=\f{1}{\sqrt{\det\lt(2\pi C\rt)}}\exp\lt(-\f{1}{2}\delta t^{\rm T}C^{-1}\delta t\rt), \n
\end{align}
where $C=N+TBT^{\rm T}$, with $N$ being the covariance matrix of $n$, $T=[M,F]$, and $B=\text{diag}(\infty,\phi)$. Here, $\infty$ is a diagonal matrix of infinities, implying that the priors for the parameters in $\epsilon$ are assumed to be flat. 

\subsection{Data}\label{sec:bayes:data}

The experimental data that we use in this work to constrain our model parameters in the inflaton potential come from two sides: PBHs and SIGWs. On the one side, the PBH abundance obtained from our model must obey the observational constraints in the relevant mass ranges. On the other side, for SIGWs, we consult the NANOGrav 15-year data set, containing the posterior distribution data of the time delay $d(f)$ for 14 frequency bins and covering the frequency range from $2.0\times10^{-9}$ to $2.8\times10^{-8}$ Hz, which can be related to the power spectrum $S(f)$ as $S(f)=T_{\rm obs}d(f)^2$. Moreover, the energy density of the model-independent free GW spectrum $\widehat{\Omega}_{\mathrm{GW}}(f)$ can be calculated from $S(f)$ as
\begin{align}
\widehat{\Omega}_{\mathrm{GW}}(f)=\frac{8\pi^4}{H_0^2}f^5S(f). \n
\end{align}

Besides SIGWs, such a free GW spectrum can also be interpreted in other different ways. One possibility is the GW spectrum generated from SMBHBs \cite{NANOGrav:2023hvm},
\begin{align}
\Omega_{\mathrm{BHB}}(f) =\f{2\pi^{2}}{3 H_{0}^{2}}A_{\mathrm{BHB}}^{2}\lt(\f{f}{\rm yr^{-1}}\rt) ^{5-\gamma_{\mathrm{BHB}}}\,{\rm yr}^{-2}, \n
\end{align}
where $A_{\mathrm{BHB}}$ is the amplitude of intrinsic noise, with 
\begin{align}
\log_{10}A_{\rm BHB}=-15.6^{+0.7}_{-0.8} \quad (68\%~{\rm CL}), \n 
\end{align}
and $\gamma_{\mathrm{BHB}}$ is the spectral index. Astrophysical and cosmological predictions suggest that the orbital evolution of binary star system driven solely by the GW emission follows 
\begin{align}
\gamma_{\rm{BHB}}=\f{13}{3}. \n 
\end{align}
However, when $\gamma_{\mathrm{BHB}}$ is treated as a free parameter, its posterior distribution can be broader. Thus, in Sec. \ref{sec:Constraints:SMBHBs}, both $A_{\rm BHB}$ and $\gamma_{\rm BHB}$ will also be considered as free model parameters in our Bayesian analysis.

\subsection{Prior distributions and Bayes factor} \label{sec:bayes:prior}

In this work, unless otherwise specified, the prior distributions of the five model parameters $\phi_0$, $\sigma$, $A_0$ (for the inflaton model), $A_{\rm BHB}$ and $\gamma_{\rm BHB}$ (for the SMBHBs) in the KKLT model are listed in Table \ref{table.1} (those in the Starobinsky model are shown in App. \ref{sec:app}). The priors of $\phi_0$, $\sigma$, and $A_0$ are chosen as uniform distributions, and the priors of $\log_{10}A_{\rm BHB}$ and $\gamma_{\rm BHB}$ are the bivariate normal distribution, with the mean ${\bm \mu}_{\mathrm{BHB}}$ and covariance matrix ${\bm \sigma}_{\mathrm{BHB}}$ being
\begin{align}
{\bm \mu}_{\mathrm{BHB}}&=\binom{-15.6}{4.7}, \quad {\bm \sigma}_{\mathrm{BHB}}=\left(\begin{array}{cc}0.28 & -0.0026 \\-0.0026 & 0.12\end{array}\right). \n
\end{align}
\begin{table}[htb]
\renewcommand\arraystretch{1.25}
\centering
\begin{tabular}{m{3.6cm}<{\centering}|m{5.0cm}<{\centering}}
\hline\hline
Parameters & Priors \\
\hline
$\phi_0$ & uniform in $[2.25,2.50]$ \\
\hline
$\sigma$ & uniform in $[0.020,0.027]$ \\
\hline
$A_0$ & uniform in $[-10^{-3},10^{-3}]$ \\
\hline
$(\log_{10}A_{\rm BHB},\gamma_{\rm BHB})$ & normal$({\bm \mu}_{\rm BHB},{\bm \sigma}_{\rm BHB})$ \\
\hline\hline
\end{tabular}
\caption{The prior distributions of the five model parameters $\phi_0$, $\sigma$, $A_0$, $A_{\rm BHB}$, and $\gamma_{\rm BHB}$. The units of $\phi_0$ and $\sigma$ (i.e., the reduced Planck mass $m_{\rm{P}}$) are omitted for simplicity.} \label{table.1}
\end{table}

When comparing the preference of different models by the observational data, we usually calculate the Bayes factor $\mathcal{B}_{12}$, defined as \cite{Thrane:2018qnx}
\begin{align}
\mathcal{B}_{12}=\frac{\mathcal{Z}_1}{\mathcal{Z}_2}, \n
\end{align}
where the subscripts $1$ and $2$ correspond to two models, and $\mathcal{Z}$ is the evidence of the relevant model, 
\begin{align}
\mathcal{Z}=\int\mathcal{L}(d|\theta)\pi(\theta)\,{\rm d}\theta,\n
\end{align}
where $d$ is the data, $\theta$ is the model parameters, $\mathcal{L}(d|\theta)$ is the likelihood function, and $\pi(\theta)$ is the prior distribution, respectively. 

\section{Constraints on the inflaton potential in the KKLT model} \label{sec:Constraints}

By combining the SIGW spectrum $\Omega_{\mathrm{GW}}(f)$ obtained from our inflaton potential and the NANOGrav 15-year data set, we can constrain the model parameters. In this work, we use the \texttt{PTArcade} code, which takes $\Omega_{\mathrm{GW}}(f)$ as input, samples the parameter space via the Markov Chain Monte Carlo method, and outputs the posterior distributions of the model parameters.

Our work consists of three steps in Secs. \ref{sec:Constraints:TwoParams}--\ref{sec:Constraints:SMBHBs}. In the first two steps, we regard the SIGWs as the unique source of the SGWB. Their difference is whether the plateau in the USR region is perfect or not, corresponding to two or three parameters in the inflaton potential. Moreover, the PBH abundance is taken into account in Sec. \ref{sec:Constraints:ThreeParams} to further reduce the allowed parameter space. Then, we also include the GWs from the SMBHBs and combine them with the SIGWs to understand the NANOGrav results. In this case, we will deal with five model parameters. Altogether, we wish to provide a thorough analysis of the NANOGrav 15-year data set with various possible interpretations.

Below, for clarity and conciseness, we focus on the KKLT model. All the analysis, figures, and discussions in the Starobinsky model are summarized in App. \ref{sec:app}, and we will find that the basic results and conclusions in these two models are very similar, supporting the generality of our work.

\subsection{Case with two model parameters} \label{sec:Constraints:TwoParams}

To exhibit our calculation procedure and obtain preliminary insights for reference below, we start from the Bayesian analysis with only two model parameters: the position $\phi_0$ and the width $\sigma$ of the perturbation $\delta V (\phi)$ on the background inflaton potential $V_{\rm b}(\phi)$. This means that the third parameter $A_0$ is fixed to be zero, in order to create a perfect plateau on $V_{\rm b}(\phi)$. The prior distributions of $\phi_0$ and $\sigma$ are listed in Table \ref{table.1}, but for better illustration, we will use $\log_{10}\sigma$ as the coordinate axes in all following figures, so $\log_{10}\sigma\in[-1.699,-1.569]$.

\begin{figure}[h]
\begin{center}
\includegraphics[width=0.6\textwidth]{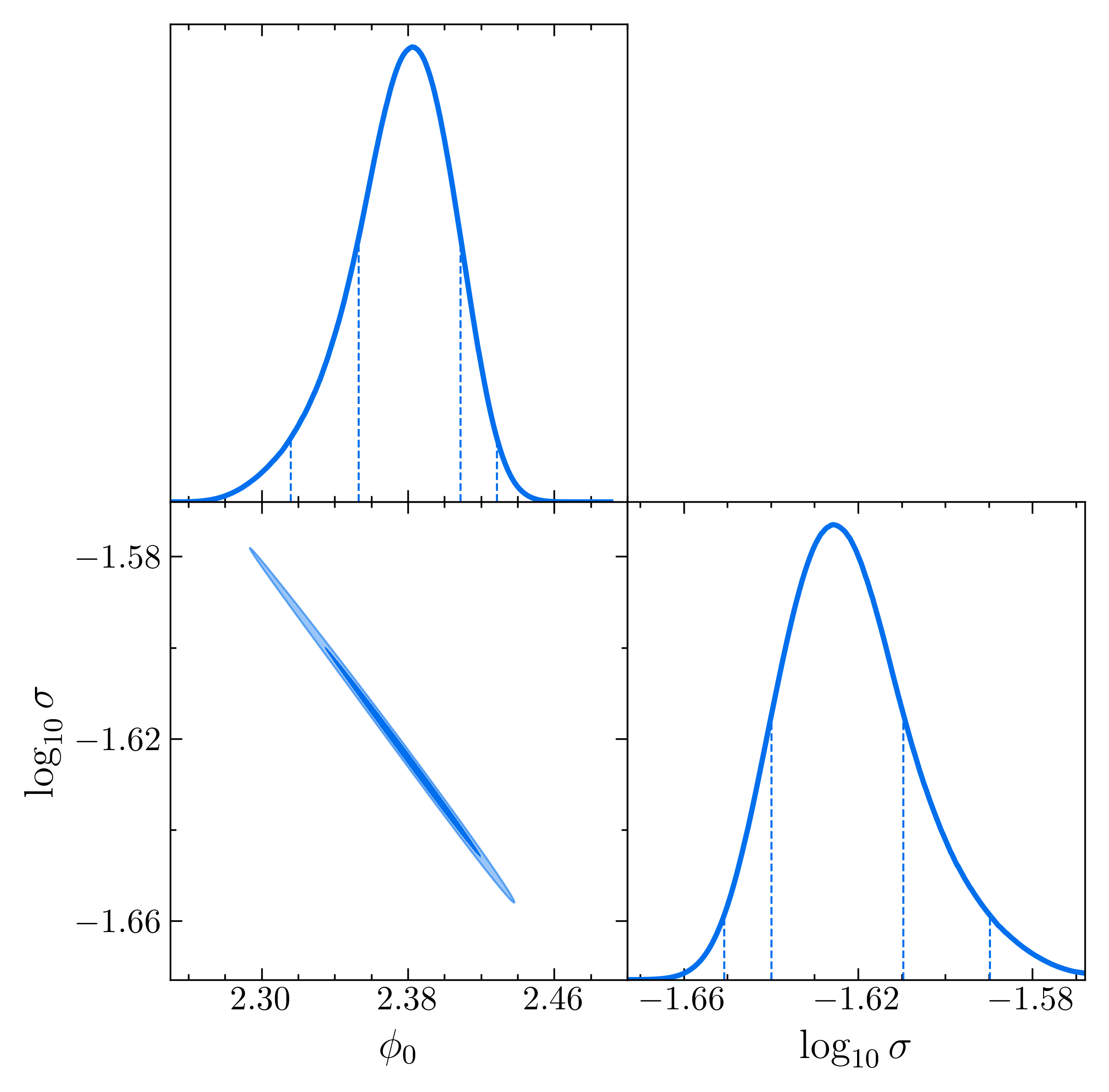}
\end{center}
\caption{The posterior distributions of the two model parameters $\phi_0$ and $\sigma$ (both in the units of $m_{\rm{P}}$). The blue shaded areas correspond to the $68\%$ and $95\%$ CLs. At the $68\%$ CL, we have $\phi_0=2.382^{+0.027}_{-0.029}$ and $\log_{10}\sigma=-1.626^{+0.016}_{-0.014}$. It can be clearly seen that $\phi_0$ and $\sigma$ exhibit a negative correlation, with strong parameter degeneracy between them.} \label{fig1}
\end{figure}

The posterior distributions of $\phi_0$ and $\sigma$ are displayed in Fig. \ref{fig1}. At the $68\%$ CL, we have
\begin{align}
\phi_0&=2.382^{+0.027}_{-0.029}, \quad \log_{10}\sigma=-1.626^{+0.016}_{-0.014}. \label{2phi2sigma}
\end{align}
Moreover, we clearly see a negative correlation between $\phi_0$ and $\sigma$, with strong parameter degeneracy in their joint posterior distribution. This makes intuitive sense. With $\sigma$ held constant, an increase of $\phi_0$ not only shifts the peak position of the SIGW spectrum $\Omega_{\rm GW}(f)$, but also increases its amplitude. Consequently, $\sigma$ must be decreased to ensure that $\Omega_{\rm GW}(f)$ satisfies the observational constraints. 

\subsection{Case with three model parameters} \label{sec:Constraints:ThreeParams}

\begin{figure}[h]
\begin{center}
\includegraphics[width=0.75\textwidth]{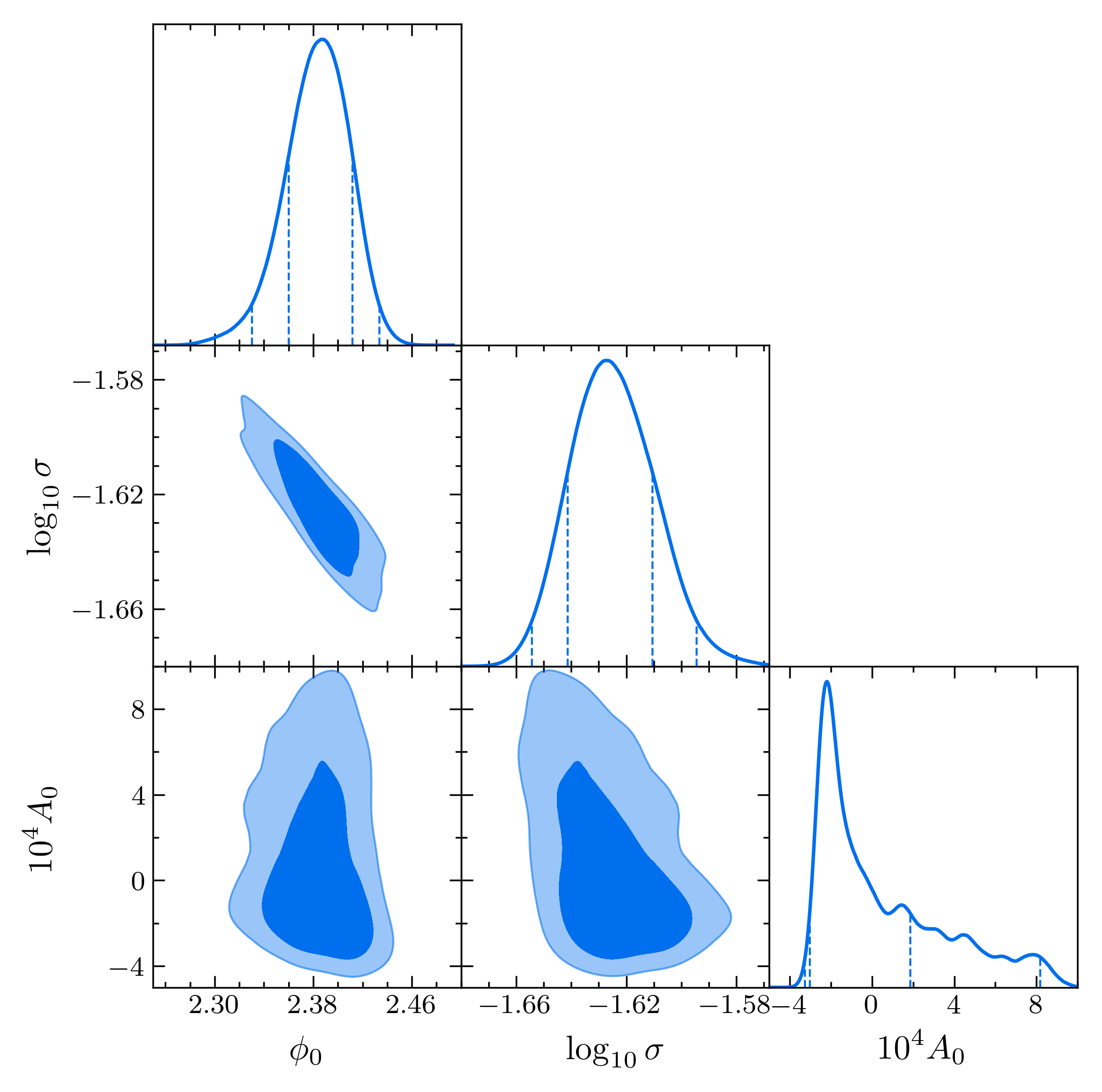}
\end{center}
\caption{Same as Fig. \ref{fig1}, but with three model parameters $\phi_0$, $\sigma$, and $A_0$. At the $68\%$ CL, we obtain $\phi_0=2.387^{+0.025}_{-0.027}$, $\log_{10}\sigma=-1.627^{+0.017}_{-0.014}$, and $A_0=-2.197^{+4.054}_{-0.841}\times10^{-4}$, respectively. The negative correlation between $\phi_0$ and $\sigma$ in their joint posterior distribution is somewhat broken by the introduction of $A_0$. In addition, $A_0$ has high probability with relatively small negative values. This observation prefers the inflation model with a small positively sloped plateau in the USR region.} \label{fig2}
\end{figure}

Now, we extend our Bayesian analysis by taking the third model parameter $A_0$ into account. This means that the plateau on $V_{\rm b}(\phi)$ is allowed to be inclined slightly. The prior distributions of $\phi_0$, $\sigma$, and $A_0$ can be found in Table \ref{table.1}, and we show our results in Fig. \ref{fig2}. At the $68\%$ CL, we obtain the posterior distributions of $\phi_0$ and $\sigma$ as
\begin{align}
\phi_0=2.387^{+0.025}_{-0.027},\quad \log_{10}\sigma=-1.627^{+0.017}_{-0.014}. \label{3phi3sigma}
\end{align}
These results remain basically the same as those in Eq. (\ref{2phi2sigma}), but their joint posterior distribution has become broader compared with Fig. \ref{fig1}, indicating that the negative correlation between $\phi_0$ and $\sigma$ can be broken to some extent by the introduction of the third parameter $A_0$. Moreover, at the $68\%$ CL, we obtain
\begin{align}
A_0=-2.197^{+4.054}_{-0.841}\times10^{-4}. \label{3A}
\end{align}
The posterior distribution of $A_0$ peaks at a relatively small negative value of $A_0=-2.197\times10^{-4}$, suggesting a positive slope of the plateau in the USR region. This result is physically reasonable and can be understood from two aspects. First, if the slope is too positive, the inflaton rolls down its potential too quickly, violating the USR condition. Second, if the slope is too negative, it is rather difficult for the inflaton to pass through the USR region. Therefore, we have adopted a narrow prior range for $A_0$, in order to avoid such situations.

\begin{figure}[h]
\begin{center}
\includegraphics[width=0.7\textwidth] {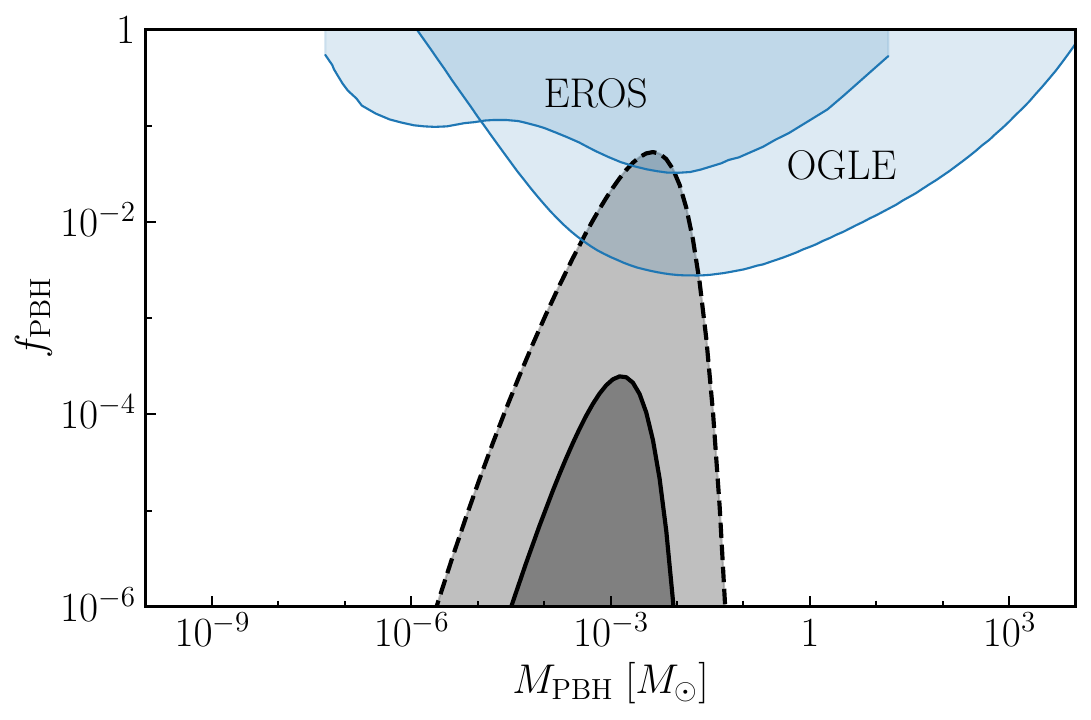}
\end{center}
\caption{The PBH abundance $f_{\rm PBH}$ obtained with two different sets of model parameters. 
The solid black line with $\phi_0=2.39$, $\log_{10}\sigma=-1.631$, and $\log_{10}A_0=-4.6$ gives reasonable $f_{\rm PBH}$, but the dashed black line with $\phi_0=2.41$, $\log_{10}\sigma=-1.641$, and $\log_{10}A_0=-4.5$ results in too large $f_{\rm PBH}$, obviously in contradiction with the constraints from Exp\'erience pour la Recherche d'Objets Sombres (EROS) \cite{Croon:2020wpr} and Optical Gravitational Lensing Experiment (OGLE) \cite{Niikura:2019kqi}. This observation indicates that the majority of the parameter spaces in Fig. \ref{fig2} leads to an overproduction of PBHs.} \label{fig3}
\end{figure}

Till now, we have merely considered the posterior distributions of the model parameters constrained by the NANOGrav 15-year data set. However, these parameters should not only guarantee the GW spectrum, but also must satisfy the constraints on the PBH abundance $f_{\rm PBH}$ at the same time. As a consequence, the parameter spaces in Fig. \ref{fig2} are still too large. To address this, we calculate $f_{\rm PBH}$ with two different sets of model parameters, both satisfying the posterior distributions in Fig. \ref{fig2}. The first set is $\phi_0=2.39$, $\log_{10}\sigma=-1.631$, and $\log_{10}A_0=-4.6$, and the second set is $\phi_0=2.41$, $\log_{10}\sigma=-1.641$, and $\log_{10}A_0=-4.5$, with their comparison illustrated in Fig. \ref{fig3}. We find that the relevant PBH masses are both around $10^{-3}~M_\odot$, where the observational constraints mainly come from the lensing experiments \cite{Niikura:2019kqi, Croon:2020wpr}. It is obvious that the first set produces reasonable $f_{\rm PBH}$ (the solid black line), but the second set leads to strong contradiction with the experimental data (the dashed black line). Actually, we will discover that the majority of the parameter spaces in Fig. \ref{fig2} results in an overproduction of PBHs, so they must be further constrained by $f_{\rm PBH}$.\footnote{Here, we should mention that there would be a risk of PBH overproduction, if the SGWB around nHz frequencies seen by PTAs is attributed to the SIGWs from the Gaussian curvature perturbation \cite{Franciolini:2023pbf}. To alleviate this problem, various attempts have been presented. The most common consideration is the introduction of non-Gaussianities \cite{2103.01056, 2211.01728} (e.g., a negative non-Gaussianity \cite{2307.00572, 2307.01102, 2310.11034, 2409.18983} or a logarithmic non-Gaussianity \cite{2411.07647}), but with some opposite opinions in Refs. \cite{2306.17834, 2412.19631}. Other investigations can also be found in the literature, with the relevant topics on the domain wall networks \cite{Gouttenoire:2023ftk}, the scalar fluctuation propagation speed \cite{Balaji:2023ehk}, the stiffer equation of state for the cosmic medium in the early Universe \cite{2309.00228, Domenech:2024rks}, the axion--gauge dynamics \cite{2307.02322}, the power spectrum with finite-width \cite{2501.00295}, and the more complex inflation models (e.g., the curvaton scenario \cite{2305.13382, 2307.03078}, the inflation model with an additional spectator tensor field \cite{2307.13109}, the Galileon inflation model \cite{2401.10925}, the radiative hybrid inflation model \cite{2402.06613}, and the bounce model \cite{2407.18976}).}

\begin{figure}[h]
\begin{center}
\includegraphics[width=0.75\textwidth]{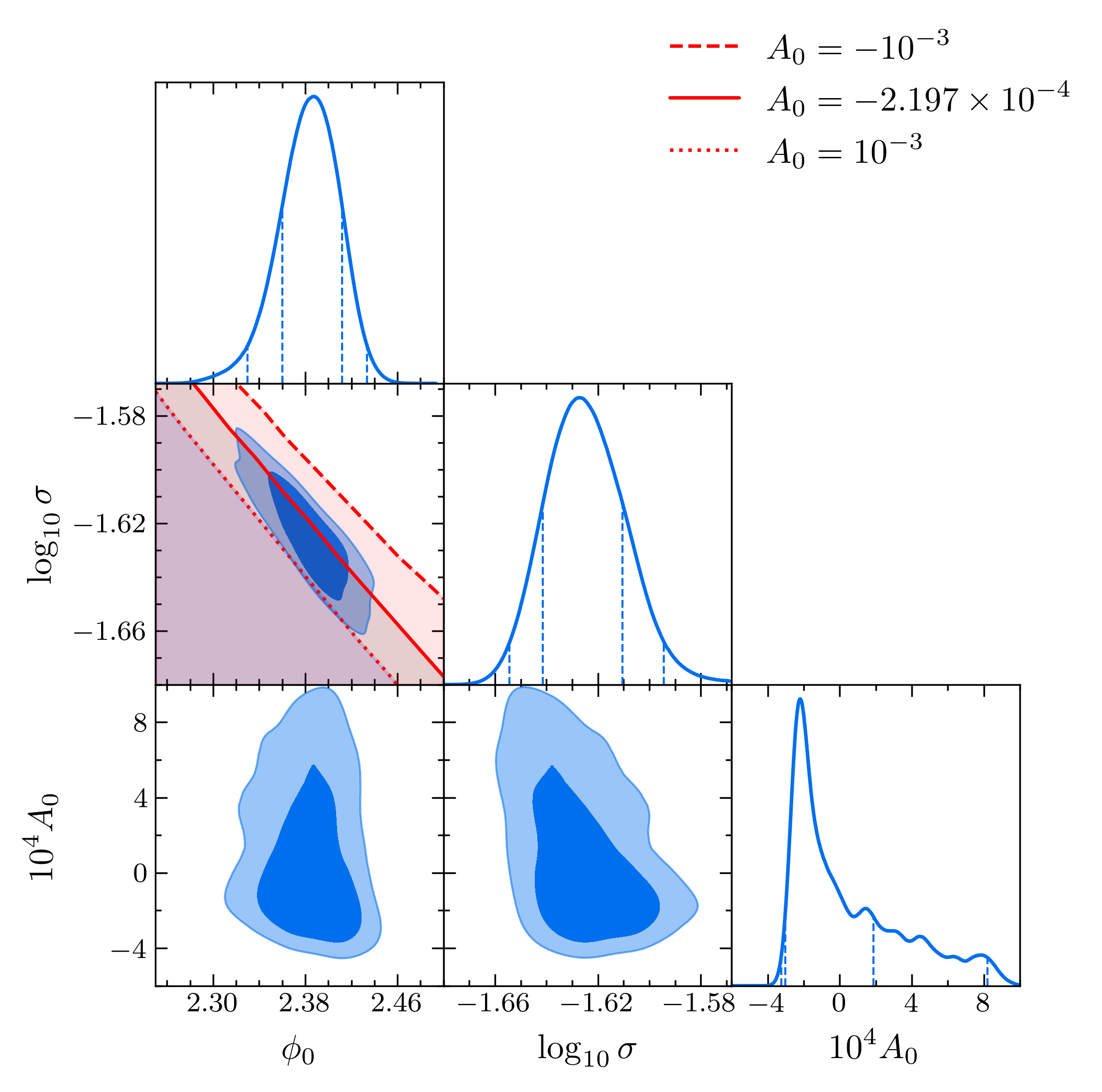} 
\end{center}
\caption{Same as Fig. \ref{fig2}, but with the constraints from the PBH abundance included. We take $A_0$ as various constants and plot the allowed regions of $\phi_0$ and $\sigma$ in their joint posterior distribution. The overlapped regions with the contours represent the parameter spaces allowed for both the SIGW and PBH constraints ( the solid red line for the maximum marginal posterior value $A_0=-2.197\times10^{-4}$, and the dashed and dotted red lines for $A_0 = -10^{-3}$ and $10^{-3}$, corresponding to its prior bounds). With the increase of $A_0$, the allowed region of $\sigma$ decreases (for the same $\phi_0$). This is because, when the slope at $\phi_0$ becomes negative, the width of the USR region must be small, otherwise the inflaton cannot roll down its potential. The three lines with different $A_0$ in the joint posterior distribution of $\phi_0$ and $\sigma$ are actually not straight. They seem so only because the prior ranges of $\phi_0$ and $\sigma$ are narrow.} \label{fig4}
\end{figure}

Consequently, it is essential to constrain the posterior distributions of the model parameters in order to satisfy the PBH abundance. To simplify this analysis, we take $A_0$ as various constants: $-2.197\times10^{-4}$ (the maximum marginal posterior value) and $\pm 10^{-3}$ (the bounds of the prior distribution), and show the allowed regions in the joint posterior distribution of $\phi_0$ and $\sigma$ in Fig. \ref{fig4}. We find that the larger $A_0$ is, the smaller the allowed region becomes. This observation indicates the decrease of $\sigma$ with the increase of $A_0$ (for the same $\phi_0$), meaning that if the USR region has a negative slope, the plateau on the background potential must be narrow enough, so that the inflaton can roll down its potential definitely.

\begin{figure}[h]
\begin{center}
\includegraphics[width=0.485\textwidth]{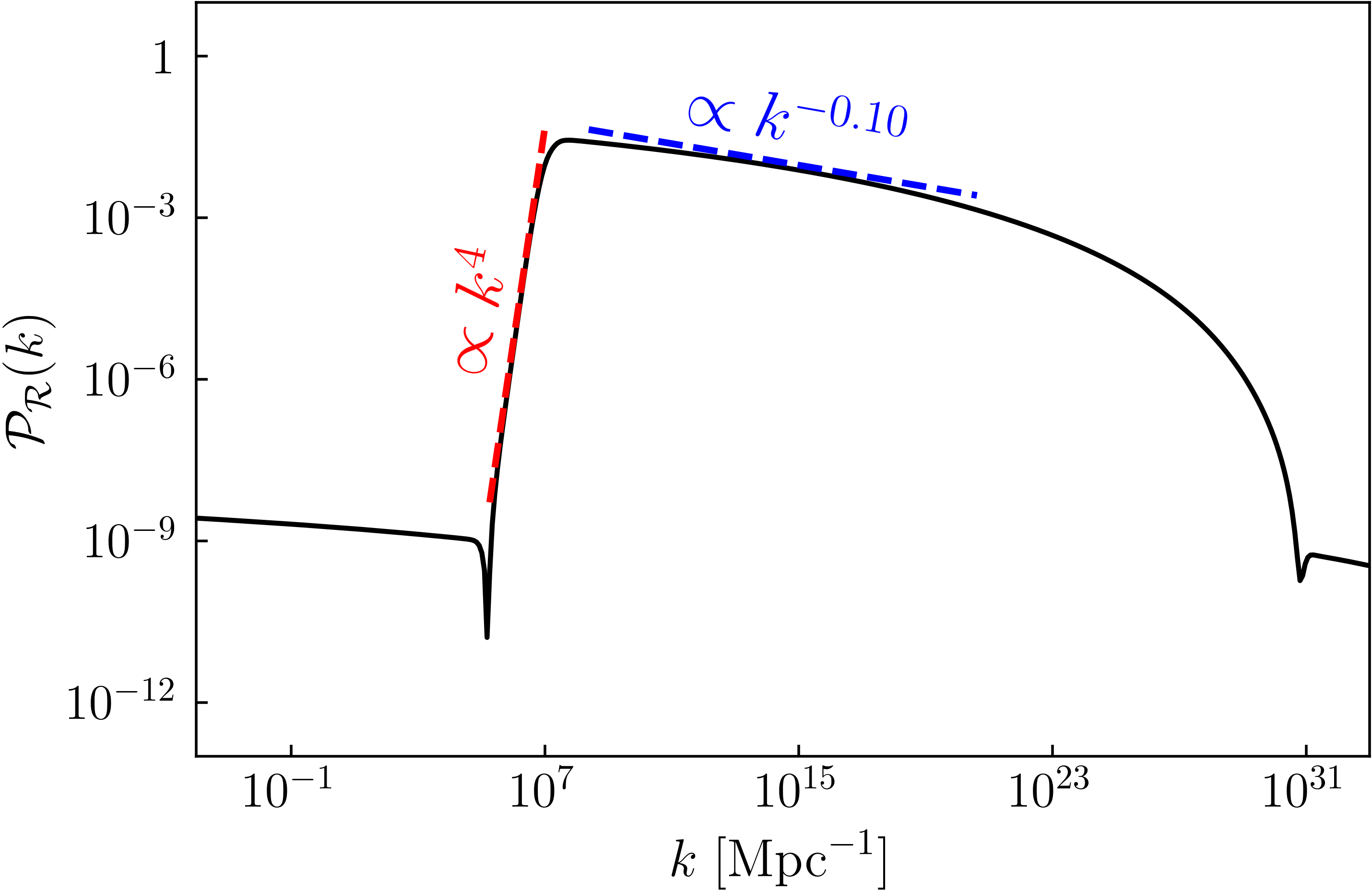} ~ 
\includegraphics[width=0.485\textwidth]{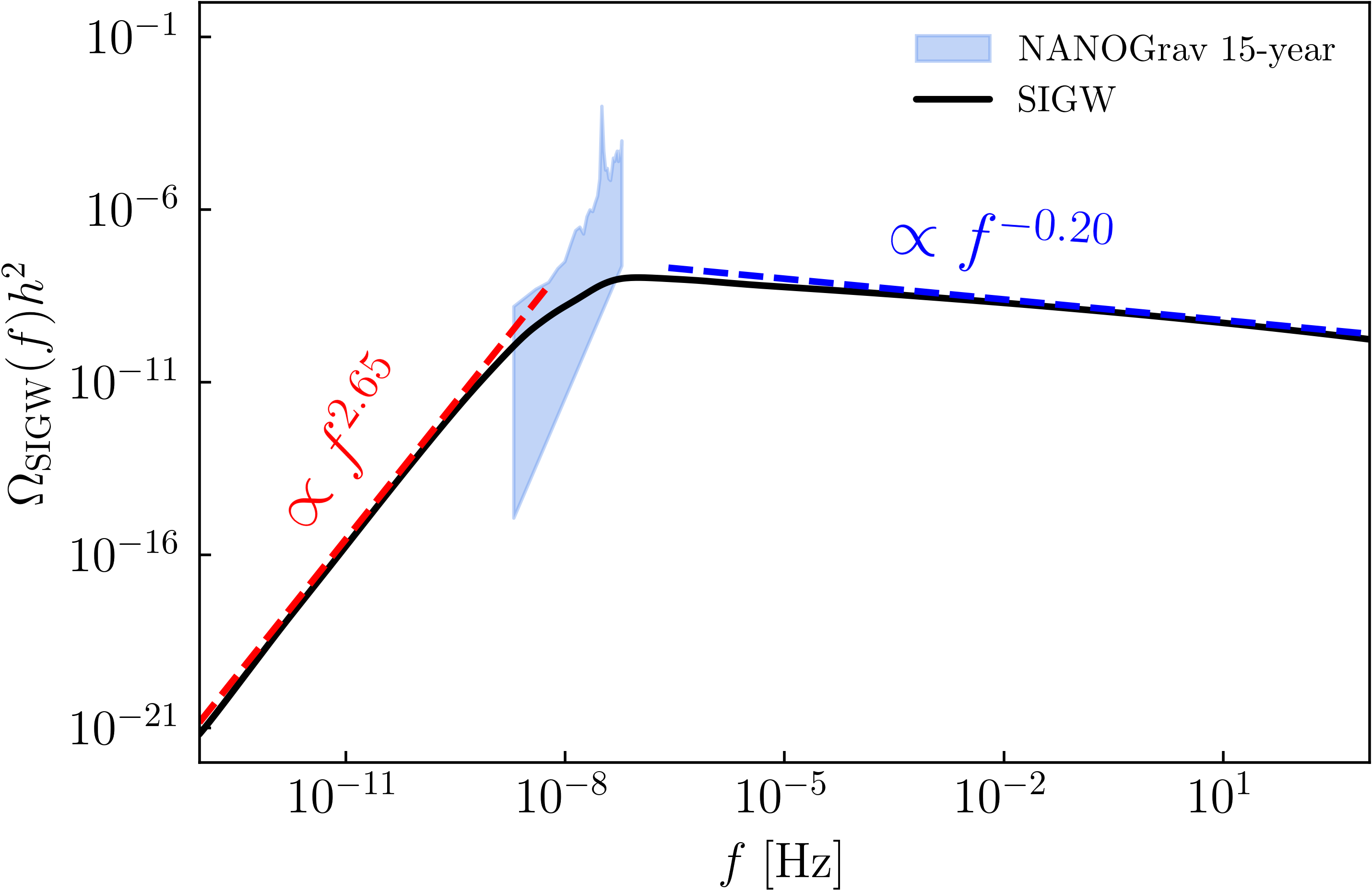}
\caption{ The scaling behaviors of the power spectrum ${\cal P}_{\cal R}(k)$ (the left panel) and the SIGW spectrum $\Omega_{\rm SIGW}(f)$ (the right panel). The model parameters read $\phi_0=2.388$, $\log_{10}\sigma=-1.623$, and $A_0=-2.197\times10^{-4}$, so as to satisfy both the PBH and SIGW constraints. ${\cal P}_{\cal R}(k)$ scales as $k^4$ in the steepest growth, then decreases as $k^{-0.10}$, and finally drops on much smaller scales. Meanwhile, $\Omega_{\rm SIGW}(f)$ scales as $f^{2.65}$ and $f^{-0.20}$ in the infrared and ultraviolet regimes.} \label{fig:ps and gw}
\end{center}
\end{figure}

Below, we choose a specific set of model parameters: $\phi_0=2.388$, $\log_{10}\sigma=-1.623$, and $A_0=-2.197\times10^{-4}$ as an example, which satisfy both the PBH and SIGW constraints, in order to discuss the scaling behaviors of the power spectrum ${\cal P}_{\cal R}(k)$ and the SIGW spectrum $\Omega_{\rm SIGW}(f)$. In Fig. \ref{fig:ps and gw}, it can be clearly seen that ${\cal P}_{\cal R}(k)$ exhibits the notable $k^4$ scaling behavior in the steepest growth, consistent with both the numerical \cite{Byrnes:2018txb} and analytical results \cite{Liu:2020oqe}, then decreases approximately as $k^{-0.10}$, and finally drops on much smaller scales.\footnote{Generally speaking, the scaling behavior of ${\cal P}_{\cal R}(k)$ on small scales depends on specific inflationary models. For more detailed discussions, see Refs. \cite{Zhao:2023zbg, Zhao:2024yzg}.} Moreover, the frequency dependence of $\Omega_{\rm SIGW}(f)$ is $f^{2.65}$ in the infrared regime and $f^{-0.20}$ in the ultraviolet regime. At even lower frequencies ($\ll$ nHz), the $f^3$ dependence of $\Omega_{\rm SIGW}(f)$ \cite{Yuan:2021qgz} is expected, but we do not show it here for simplicity.

\begin{figure}[h]
\begin{center}
\includegraphics[width=0.75\textwidth]{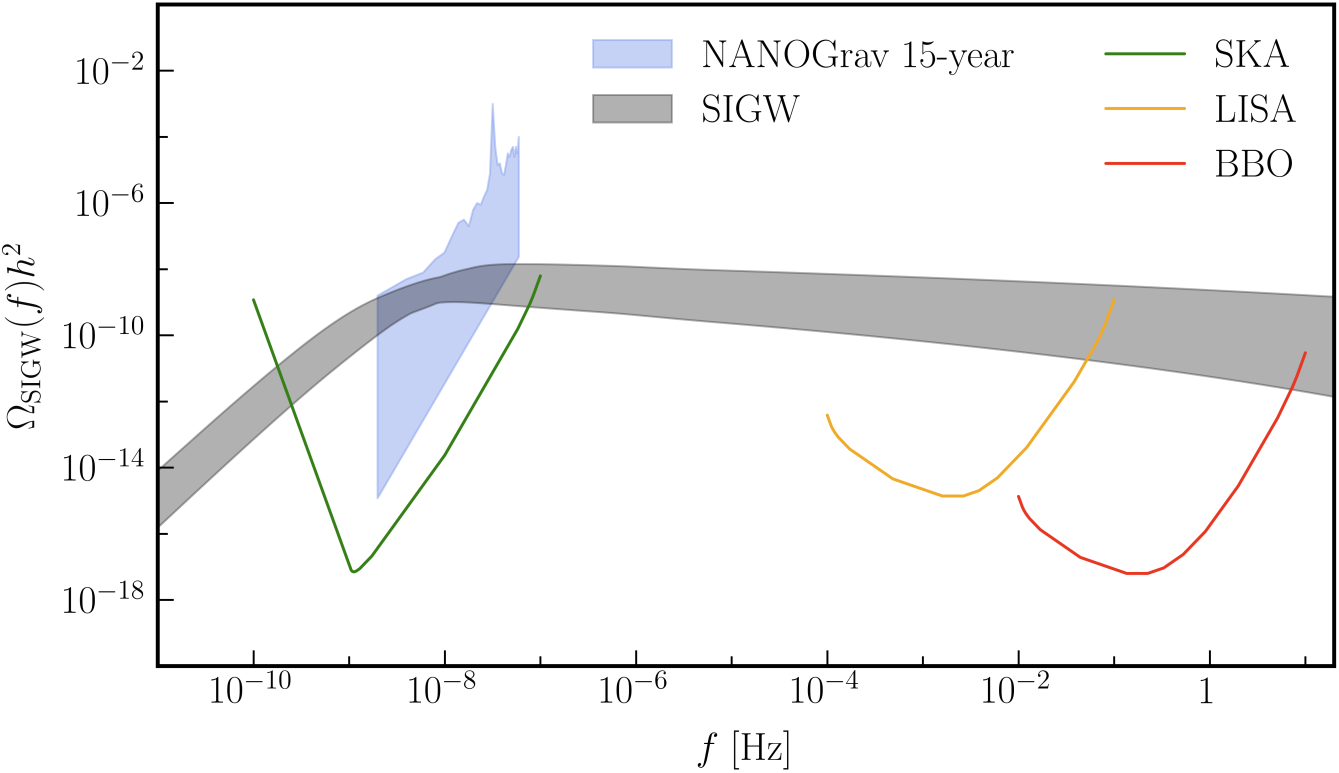}
\end{center}
\caption{The envelope of the SIGW spectra $\Omega_{\mathrm{GW}}(f)$ (the grey belt) obtained from the model parameters satisfying both the constraints from the NANOGrav 15-year data set and the PBH abundance, with $A_0 = -2.197\times10^{-4}$. The shaded light blue region represents the periodogram for the free spectral process from the NANOGrav results, and our SIGW spectra pass through it with appropriate amplitudes and spectral indices. In addition, as the SIGW spectra are already above the sensitivity curves of three next-generation GW detectors: SKA, LISA, and BBO, the SIGWs are expected to be observed in future.} \label{fig5}
\end{figure}

Furthermore, by means of a modified version of the \texttt{SIGWfast} code \cite{Witkowski:2022mtg}, the envelope of the SIGW spectrum is plotted in Fig. \ref{fig5}, satisfying the PBH abundance constraint, with $A_0=-2.197\times10^{-4}$ and $\phi_0$ and $\sigma$ in the allowed region in their joint posterior distribution in Fig. \ref{fig4}. Thus, Fig. \ref{fig:ps and gw} can be regarded as a special case in Fig. \ref{fig5}. For comparison, we also include the sensitivity curves of three next-generation GW detectors: the Square Kilometer Array (SKA) \cite{Janssen:2014dka}, Laser Interferometer Space Antenna (LISA) \cite{LISA:2017pwj}, and Big Bang Observer (BBO) \cite{Corbin:2005ny}. We observe that the SIGW spectrum obtained from our inflation model (the grey belt) not only successfully accounts for the NANOGrav 15-year data set with appropriate amplitude and spectral index, but is also detectable by the GW detectors in future. It should be mentioned that the currently observed signal in the PTA band may limit the sensitivities of the future detectors \cite{Babak:2024yhu}, but we will not consider it in this work.


\subsection{Case with SMBHBs} \label{sec:Constraints:SMBHBs}

\begin{figure}[h]
\begin{center}
\includegraphics[width=0.75\textwidth]{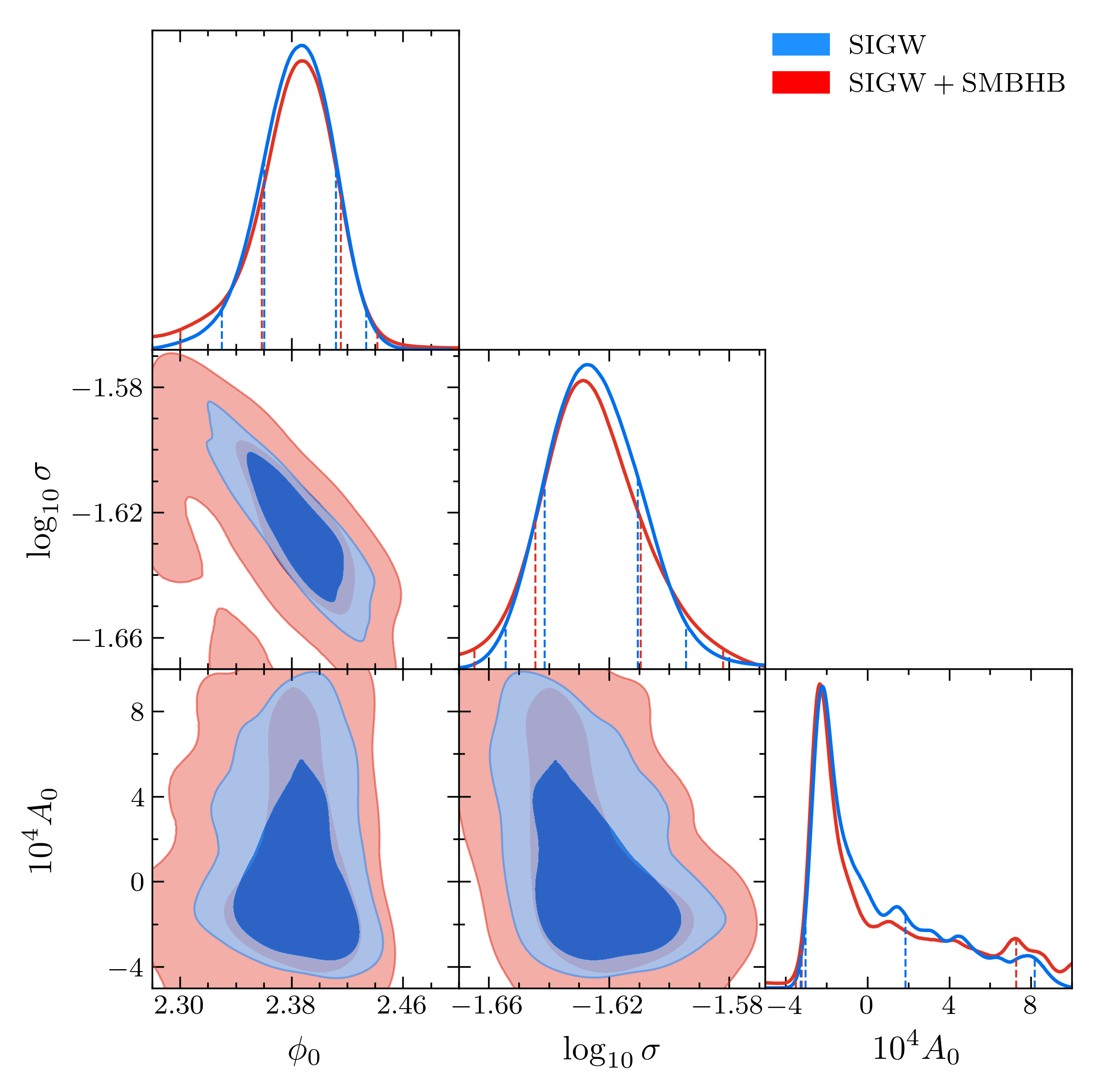} 
\end{center}
\caption{Same as Fig. \ref{fig2}, but with the impact of SMBHBs taken into account. The posterior distributions at the $68\%$ CL read $\phi_0=2.389_{-0.030}^{+0.026}$, $\log_{10}\sigma=-1.629_{-0.015}^{+0.020}$, and $A_0=-2.334_{-0.882}^{+9.613}\times10^{-4}$, respectively. We plot the three-parameter SIGW case and the five-parameter $\rm SIGW+SMBHB$ case together for comparison. The joint posterior distributions in the latter are broader. This is because, when only SIGW is considered, the GW spectrum from some parameter combinations is too low to account for the experimental data. However, if the contribution from SMBHBs is large enough, these combinations become allowed again.} \label{fig6} 
\end{figure}

\begin{figure}[h]
\begin{center}
\includegraphics[width=1.0\textwidth]{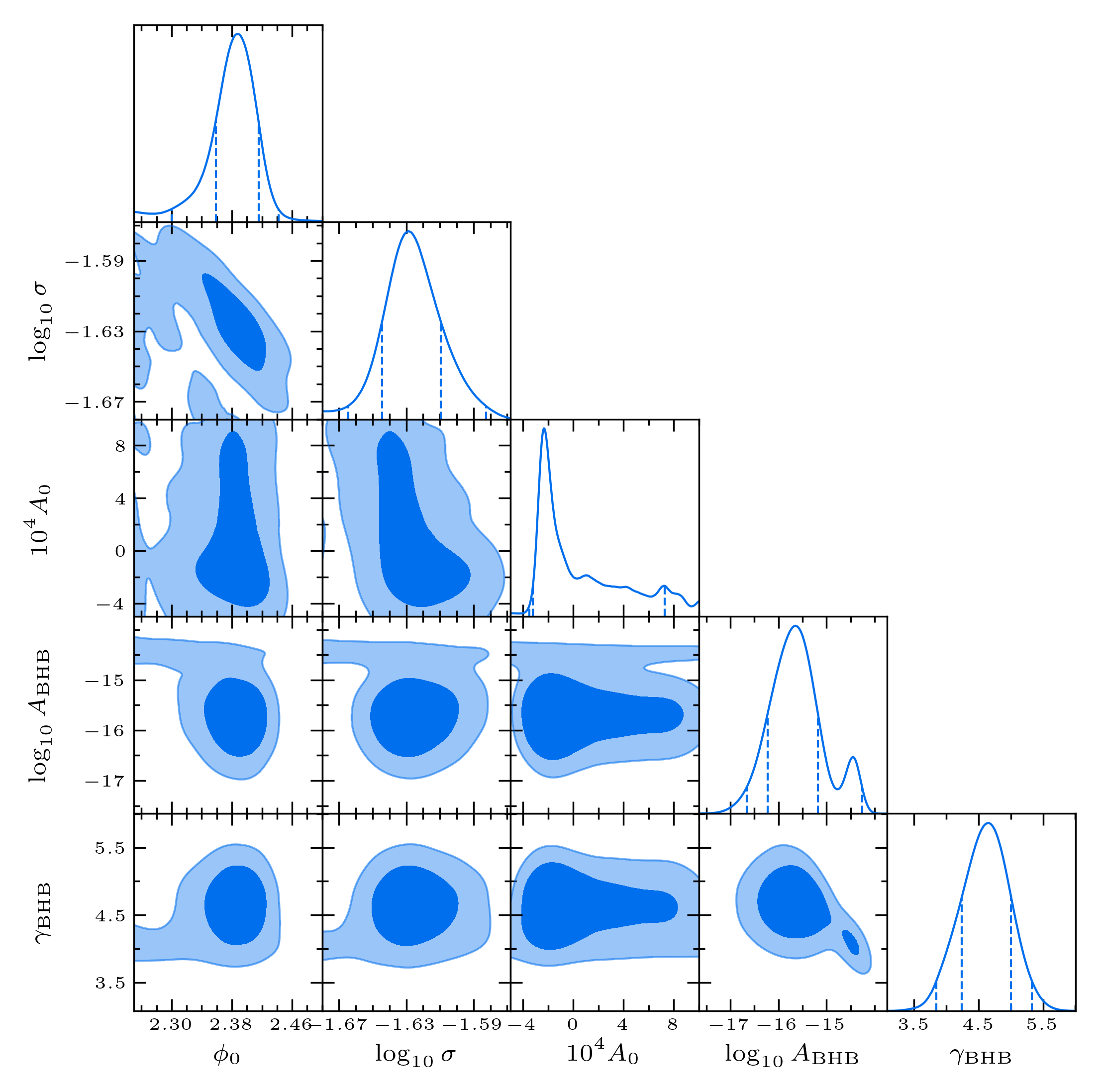} 
\end{center}
\caption{The posterior distributions of the model parameters with the impact of SMBHBs included. There are five model parameters altogether: $\phi_0$, $\sigma$, $A_0$ (for the inflation model), $A_{\mathrm{BHB}}$, and $\gamma_{\mathrm{BHB}}$ (for the SMBHBs). The posterior distribution of $A_{\rm BHB}$ has two peaks, reflecting the competition between SIGWs and SMBHBs. Different peak corresponds to different component that dominates the GW spectrum, consistent with the analysis by the NANOGrav collaboration in Ref. \cite{NANOGrav:2023hvm}. The value of the spectral index $\gamma_{\rm BHB}=13/3$ is within its $68\%$ CL when treated as a free parameter.} \label{fig7} 
\end{figure}

Last, we consider the most general case with the impact of SMBHBs taken into account. SMBHBs can form during the hierarchical merging process of galaxies in structure formation, and their relevant SGWB lies in the PTA band \cite{Rajagopal:1994zj, Jaffe:2002rt, Burke-Spolaor:2018bvk}. Therefore, we will explore the scenario, in which the SGWB is composed of the SIGWs and the GWs emitted from SMBHBs together, so the total GW spectrum is
\begin{align}
\Omega_{\mathrm{GW}} = \Omega_{\mathrm{SIGW}}+\Omega_{\mathrm{BHB}}. \n
\end{align}
Now, we face the posterior distributions of five parameters: $\phi_0$, $\sigma$, $A_0$ (for the inflation model), $A_{\rm BHB}$, and $\gamma_{\rm BHB}$ (for the SMBHBs). Our results are presented in Figs. \ref{fig6} and \ref{fig7}. At the $68\%$ CL, we have
\begin{align}
\phi_0&= 2.389_{-0.030}^{+0.026}, \label{5phi}\\
\log_{10}\sigma&=-1.629_{-0.015}^{+0.020}, \label{5sigma}\\
A_0&=-2.334_{-0.882}^{+9.613}\times10^{-4}, \label{5A}\\
\log_{10}A_{\rm BHB}&=-15.646_{-0.557}^{+0.467}, \label{5ABHB}\\
\gamma_{\rm BHB}&=4.653_{-0.414}^{+0.351}. \label{5gamma}
\end{align}

From Eqs. \eqref{5phi}--\eqref{5A} and Fig. \ref{fig6}, we observe that all the joint posterior distributions of $\phi_0$, $\sigma$, and $A_0$ become broader than those in the three-parameter case. This is because, when only SIGW is considered, the GW spectrum from some parameter combinations is too low to match the experimental data. However, as long as the contribution from SMBHBs is large enough, these parameter combinations again become allowed (e.g., within the $95\%$ CL). 

Furthermore, from Fig. \ref{fig7}, we notice that the posterior distribution of $A_{\rm BHB}$ has two peaks, reflecting the competition between SIGWs and SMBHBs. Different peak corresponds to different component that dominates the GW spectrum, consistent with the relevant analysis by the NANOGrav collaboration in Ref. \cite{NANOGrav:2023hvm}. When $A_{\rm BHB}$ is small, the GW spectrum is dominated by SIGWs; when $A_{\rm BHB}$ is large, the GW spectrum is driven by SMBHBs. Also, the spectral index $\gamma_{\rm BHB}=13/3$ is within the $68\%$ CL when it is treated as a free parameter. 

\subsection{Bayes factors} \label{sec:b}

Finally, to complete our analysis, we calculate the Bayes factors between different models. With the parameters in Eqs. (\ref{2phi2sigma})--(\ref{5gamma}), the Bayes factor between the two-parameter and SMBHB-only models is 
\begin{align}
{\cal B}=\f{{\cal Z}_{\rm SIGW_2}} {{\cal Z}_{\rm SMBHB}}=20.21, \label{B23}
\end{align}
the Bayes factor between the three-parameter and SMBHB-only models is
\begin{align} 
{\cal B}=\f{{\cal Z}_{\rm SIGW_3}} {{\cal Z}_{\rm SMBHB}}=8.01, \label{B3S}
\end{align}
and the Bayes factor between the five-parameter and SMBHB-only models is
\begin{align} 
{\cal B}=\f{{\cal Z}_{\rm SIGW_3+SMBHB}} {{\cal Z}_{\rm SMBHB}}=7.36. \label{B5S}
\end{align}
 
The above Bayes factors between different models reveal a clear hierarchy: the two-parameter model $\rm SIGW_2$ is strongly favored; the three-parameter model $\rm SIGW_3$ and the five-parameter model $\rm SIGW_3 + SMBHB$ are comparable but weaker. However, the fact that the two-parameter model has the largest Bayes factor may be merely caused by fixing $A_0=0$ to realize a perfect plateau in the USR region. Since the SIGW spectrum essentially depend on three model parameters, we can only claim that the three-parameter model is slightly better than the five-parameter one, or equivalently, Bayesian analysis shows no preference with or without SMBHBs. This conclusion is comprehensible, because despite the increase of the number of free model parameters, the complex five-parameter model does not lead to a better fitting to the data. The likelihood does not change much, but meanwhile the prior space expands, so the five-parameter model is not superior to the three-parameter one. Nonetheless, all the two-, three-, and five-parameter models are significantly better than the SMBHB-only model, suggesting that the interpretation of the SGWB from the NANOGrav observations only via the GW spectrum from SMBHBs is not preferred in our inflation model.

\section{Conclusion} \label{sec:Conclusion}

The exploration on PBHs has aroused increasing interest in physics community in recent years. Besides its possibility as a natural candidate of DM and the source of BHB merger events, another research hotspot is the relevant SIGWs produced with PBHs simultaneously as a second-order effect. Since the detection of the potential SGWB in the PTA band, especially in the NANOGrav 15-year data set, SIGWs have become a promising interpretation in addition to the GWs from SMBHBs. To generate intensive SIGWs with certain amplitudes and frequencies, the usual SR inflation is not enough, and a USR phase is needed, in order to enhance the primordial power spectrum on small scales. The USR phase can be realized by imposing a perturbation on the background inflaton potential, and to determine the position and shape of this perturbation is the purpose of our work. For comparison, we verify the robustness of our conclusions against two different background inflaton potentials: the KKLT and Starobinsky models, and the similar results confirm the generality of our work.


In this paper, we construct an anti-symmetric perturbation $\delta V(\phi)$ in Eq. \eqref{VVV} with three parameters: $\phi_0$, $\sigma$, and $A$ (or equivalently, $A_0$ in Eq. \eqref{VAA}), describing the position, width, and slope of $\delta V(\phi)$, and these model parameters further determine the SIGW spectrum. In addition to the SIGWs from the USR inflation, we also consider the GW spectrum from SMBHBs, which is characterized by another two parameters: the amplitude $A_{\rm BHB}$ and spectral index $\gamma_{\rm BHB}$. Altogether, we encounter the Bayesian analysis with five model parameters, which can be performed via the \texttt{PTArcade} code, with their prior distributions summarized in Tables \ref{table.1} and \ref{table.2}.


Our work is composed of three steps. First, in order to reduce computation cost and obtain preliminary results for reference, we directly set the parameter $A_0=0$, so that the plateau in the USR region is perfect. The posterior distributions of the two parameters $\phi_0$ and $\sigma$ can be found in Eq. \eqref{2phi2sigma}, and we clearly observe the negative correlation in their joint posterior distribution in Fig. \ref{fig1}, with strong parameter degeneracy.

Then, we set $A_0$ as a free parameter, meaning that the plateau in the USR region is allowed to be slightly inclined. In this three-parameter case, the joint posterior distribution of $\phi_0$ and $\sigma$ becomes broader, and the posterior distribution of $A_0$ peaks at $-2.197\times10^{-4}$, as can be seen in Fig. \ref{fig2}. This feature indicates that a small positively sloped plateau on the background inflaton potential is preferred. Moreover, we find that some parameter spaces actually overproduce PBHs in certain mass region, as illustrated in Fig. \ref{fig3}. Thus, we further show the allowed parameter spaces in Fig. \ref{fig4}, which satisfy both the SIGW and PBH constraints. At last, we plot the SIGW spectrum in Fig. \ref{fig5} and find that the SIGWs not only account for the SGWB, but are also detectable by the next-generation GW detectors. 


Finally, we investigate the case with the impact of SMBHBs. We perform the full five-parameter Bayesian analysis, provide their posterior distributions in Figs. \ref{fig6} and \ref{fig7}, and calculate the Bayes factors between different models in Eqs. \eqref{B23}--\eqref{B5S}. The Bayes factors indicate that the NANOGrav 15-year data set show preference for the two-parameter model, but this may be an artifact of fixing $A_0=0$. Hence, we can only confidently claim that all the two-, three-, and five-parameter models are significantly favored in Bayesian analysis than the SMBHB-only model.

Altogether, the overall circumstance seems not to be improved by increasing the number of free model parameters. Nevertheless, this tendency is consistent with Ref. \cite{NANOGrav:2023hvm}, in which the NANOGrav collaboration analyzed the Bayes factors between different models with or without SMBHBs, concluding that some models perform better with considering SMBHBs, some are roughly equivalent, while others perform even worse. This primarily depends on whether the increase of model parameters provides a better fitting to the experimental data, and in our present work, it does not yield a substantially better fitting. Therefore, in summary, the simple two-parameter model with a perfect plateau in the USR region has the largest evidence, and is thus favored by the NANOGrav 15-year data set, though it may be just due to the demand that $A_0$ vanishes. Moreover, the interpretation of the potential SGWB from the NANOGrav observations via only the GWs generated by SMBHBs is not preferred in our work, both in the KKLT and Starobinsky models.

Last, we should mention that we have not discussed the non-Gaussian effects in this work. This includes both the nonlinear relation between the primordial curvature perturbation ${\cal R}$ and the density contrast $\delta$, as well as the intrinsic non-Gaussianity in ${\cal R}$. These non-Gaussianities may lead to notable shifts in the allowed parameter regions and different interpretations of the SGWB, and this will be the topic of our future studies.

\acknowledgments

We thank Yi-Zhong Fan, Lei Feng, Gabriele Franciolini, Zhiping Jin, Xiao-Hui Liu, Dominik J. Schwarz, Bing-Yu Su, Ji-Xiang Zhao, and Xue Zhou for fruitful discussions. This work is supported by the Fundamental Research Funds for the Central Universities of China (No. N232410019).

\appendix

\section{Constraints on the inflaton potential in the Starobinsky model} \label{sec:app}

In the Appendix, we show the Bayesian analysis of the three cases in the Starobinsky model. The contents below are parallel to those in Sec. \ref{sec:Constraints}. The priors of $\phi_0$, $\sigma$, and $A_0$ are chosen as uniform distributions, and the priors of $\log_{10}A_{\rm BHB}$ and $\gamma_{\rm BHB}$ are the bivariate normal distribution, as listed in Table \ref{table.2}. 
\begin{table}[htb]
\renewcommand\arraystretch{1.25}
\centering
\begin{tabular}{m{3.6cm}<{\centering}|m{5.0cm}<{\centering}}
\hline\hline
Parameters & Priors \\
\hline
$\phi_0$ & uniform in $[4.8,5.0]$ \\
\hline
$\sigma$ & uniform in $[0.017,0.022]$ \\
\hline
$A_0$ & uniform in $[-10^{-3},10^{-3}]$ \\
\hline
$(\log_{10}A_{\rm BHB},\gamma_{\rm BHB})$ & normal$({\bm \mu}_{\rm BHB},{\bm \sigma}_{\rm BHB})$ \\
\hline\hline
\end{tabular}
\caption{The prior distributions of the five model parameters $\phi_0$, $\sigma$, $A_0$, $A_{\rm BHB}$, and $\gamma_{\rm BHB}$ in the Starobinsky model. The priors of $A_0$, $A_{\rm BHB}$, and $\gamma_{\rm BHB}$ are the same as those in the KKLT model, the prior of $\sigma$ is also alike, but the prior of $\phi_0$ shifts to larger values, as the inflaton sets out from larger initial position.} \label{table.2}
\end{table}

\begin{figure}[h]
\begin{center}
\includegraphics[width=0.6\textwidth]{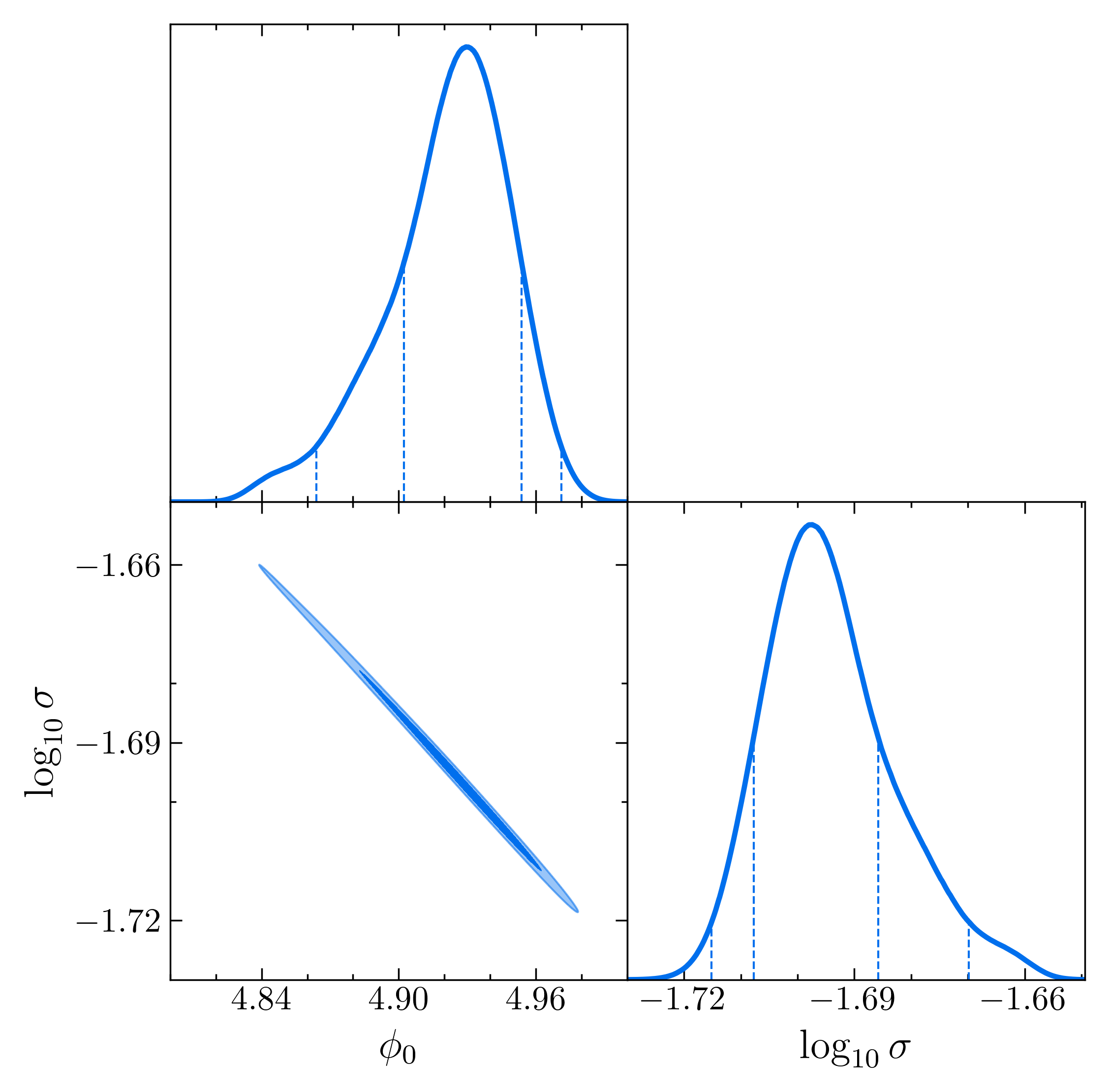}
\end{center}
\caption{The posterior distributions of the model parameters $\phi_0$ and $\sigma$ in the Starobinsky model.} \label{fig:S2}
\end{figure}

\begin{figure}[h]
\begin{center}
\includegraphics[width=0.75\textwidth]{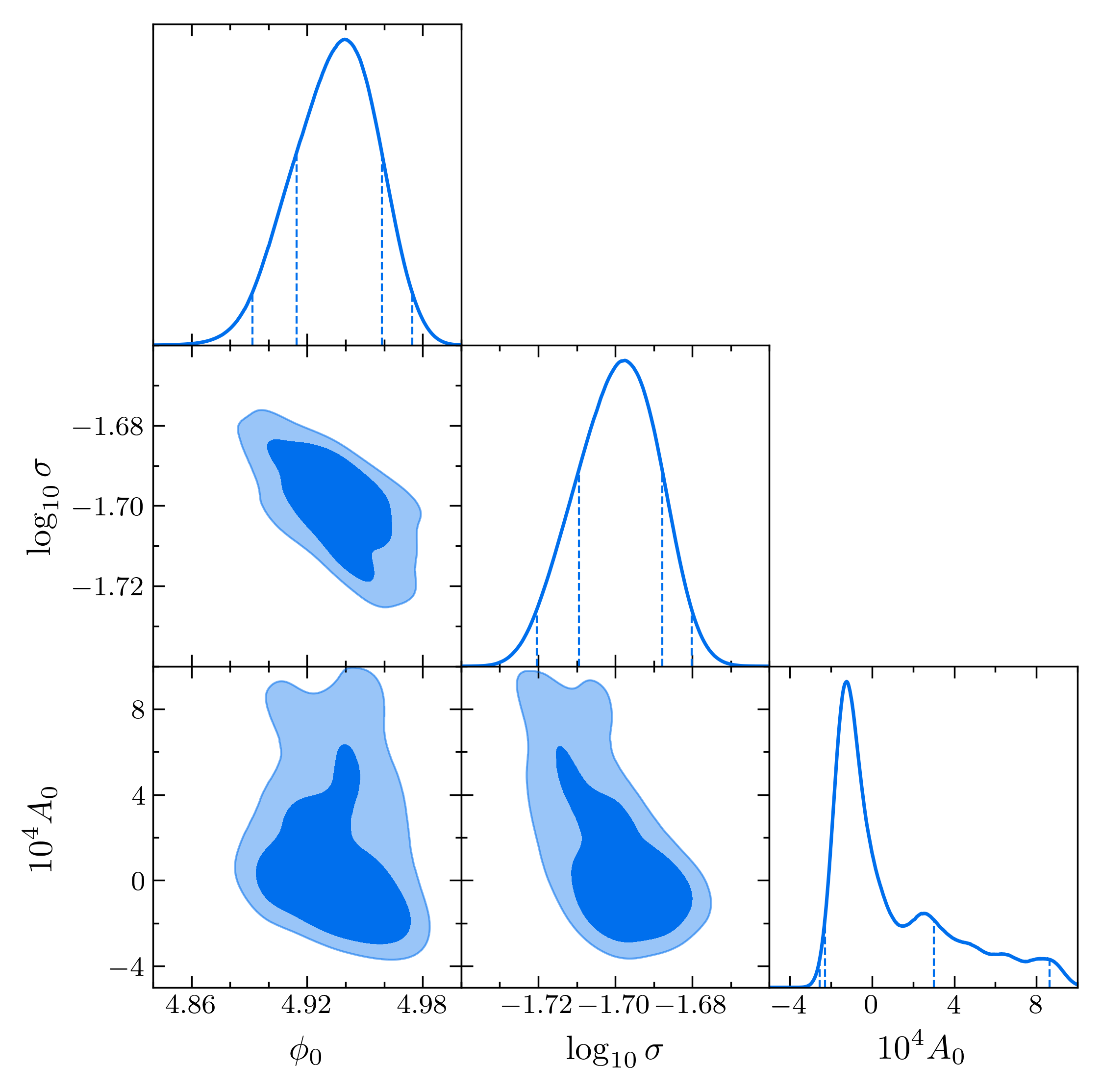}
\end{center}
\caption{The posterior distributions of the model parameters $\phi_0$, $\sigma$, and $A_0$ in the Starobinsky model.} \label{fig:S3}
\end{figure}

\begin{figure}[h]
\begin{center}
\includegraphics[width=1\textwidth]{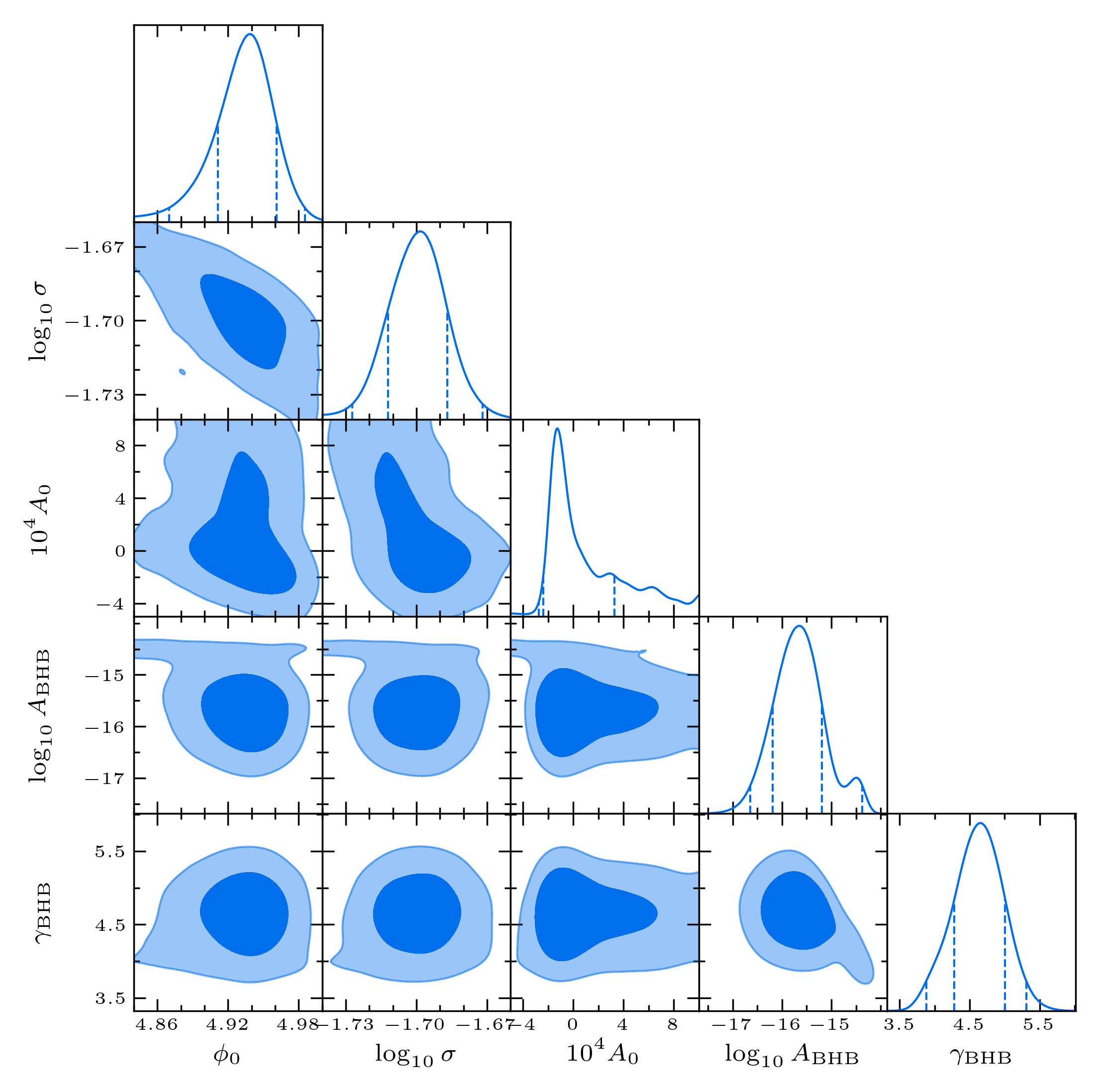}
\end{center}
\caption{The posterior distributions of the model parameters $\phi_0$, $\sigma$, $A_0$, $A_{\rm BHB}$, and $\gamma_{\rm BHB}$ in the Starobinsky model.} \label{fig:S5}
\end{figure}

First, in the case with two model parameters, the posterior distributions of $\phi_0$ and $\sigma$ are plotted in Fig. \ref{fig:S2}. At the $68\%$ CL, we have
\begin{align}
\phi_0=4.929^{+0.024}_{-0.027}, \quad \log_{10}\sigma=-1.697^{+0.011}_{-0.010}. \n 
\end{align}
Then, in the case with three model parameters, the posterior distributions of $\phi_0$, $\sigma$, and $A_0$ are shown in Fig. \ref{fig:S3}. At the $68\%$ CL, we have
\begin{align}
\phi_0=4.939^{+0.019}_{-0.024}, \quad \log_{10}\sigma=-1.697^{+0.009}_{-0.012}, \quad A_0=-1.238^{+4.250}_{-1.313}\times10^{-4}. \n 
\end{align}
Last, in the case with SMBHBs, the posterior distributions of $\phi_0$, $\sigma$, $A_0$, $A_{\rm BHB}$, and $\gamma_{\rm BHB}$ are displayed in Fig. \ref{fig:S5}. At the $68\%$ CL, we obtain
\begin{align}
\phi_0&=4.938^{+0.023}_{-0.027}, \quad \log_{10}\sigma=-1.698^{+0.011}_{-0.013}, \quad A_0=-1.280^{+4.573}_{-1.109}\times10^{-4}, \n \\
\log_{10}A_{\rm BHB}&=-15.654^{+0.465}_{-0.534} , \quad \gamma_{\rm BHB}=4.657^{+0.351}_{-0.375} . \n
\end{align}

Finally, the Bayes factor between the two-parameter and SMBHB-only models is 
\begin{align}
{\cal B}=\f{{\cal Z}_{\rm SIGW_2}} {{\cal Z}_{\rm SMBHB}}=21.49, \label{B23Sta}
\end{align}
the Bayes factor between the three-parameter and SMBHB-only models is
\begin{align} 
{\cal B}=\f{{\cal Z}_{\rm SIGW_3}} {{\cal Z}_{\rm SMBHB}}=8.26, \label{B3SSta}
\end{align}
and the Bayes factor between the five-parameter and SMBHB-only models is
\begin{align} 
{\cal B}=\f{{\cal Z}_{\rm SIGW_3+SMBHB}} {{\cal Z}_{\rm SMBHB}}=12.75. \label{B5SSta}
\end{align}

In summary, from various perspectives of the intuitive observation of the inflaton potentials, the posterior distributions of the model parameters, and the corresponding Bayes factors, the Starobinsky model exhibits significant similarity with the KKLT model. For instance, a slightly positive slope in the USR region is still preferred, and the SMBHB-only model is not favored by the data. The only difference between the two models is that their USR regions are at different positions, but this is understandable, as the potential in the Starobinsky model decreases more slowly than in the KKLT model.

\end{document}